\documentclass[12pt]{article}
\usepackage{amsfonts}
\usepackage{latexsym}
\usepackage{times}

\catcode`\@=11

\newcount\hour
\newcount\minute
\newtoks\amorpm \hour=\time\divide\hour by 60\minute
=\time{\multiply\hour by 60 \global\advance\minute by-\hour}
\edef\standardtime{{\ifnum\hour<12 \global\amorpm={am}%
        \else\global\amorpm={pm}\advance\hour by-12 \fi
        \ifnum\hour=0 \hour=12 \fi
        \number\hour:\ifnum\minute<10
        0\fi\number\minute\the\amorpm}}
\edef\militarytime{\number\hour:\ifnum\minute<10
0\fi\number\minute}

\def\draftlabel#1{{\@bsphack\if@filesw {\let\thepage\relax
   \xdef\@gtempa{\write\@auxout{\string
      \newlabel{#1}{{\@currentlabel}{\thepage}}}}}\@gtempa
   \if@nobreak \ifvmode\nobreak\fi\fi\fi\@esphack}
        \gdef\@eqnlabel{#1}}
\def\@eqnlabel{}
\def\@vacuum{}
\def\marginnote#1{}
\def\draftmarginnote#1{\marginpar{\raggedright\scriptsize\tt#1}}
\def\draft{
        \pagestyle{plain}
        \overfullrule=2pt
        \oddsidemargin -.5truein
        \def\@oddhead{\sl \phantom{\today\quad\militarytime} \hfil
        \smash{\Large\sl DRAFT} \hfil \today\quad\militarytime}
        \let\@evenhead\@oddhead
        \let\label=\draftlabel
        \let\marginnote=\draftmarginnote
        \def\ps@empty{\let\@mkboth\@gobbletwo
        \def\@oddfoot{\hfil \smash{\Large\sl DRAFT} \hfil}
        \let\@evenfoot\@oddhead}
        \def\@eqnnum{(\theequation)\rlap{\kern\marginparsep\tt\@eqnlabel}%
        \global\let\@eqnlabel\@vacuum}  }

\newcommand{\rf}[1]{(\ref{#1})}
\renewcommand{\theequation}{\thesection.\arabic{equation}}
\renewcommand{\thefootnote}{\fnsymbol{footnote}}
\newcommand{\newsection}{    
\setcounter{equation}{0}\section}
\def\appendix#1{\addtocounter{section}{1}\setcounter{equation}{0}
\renewcommand{\thesection}{\Alph{section}}
\section*{Appendix \thesection\protect\indent \parbox[t]{11.15cm}{#1}}
\addcontentsline{toc}{section}{Appendix \thesection\ \ \ #1}}

\def\im{{\rm im}}

\def\be{\begin{equation}}
\def\ee{\end{equation}}
\def\beq{\begin{eqnarray}}
\def\eeq{\end{eqnarray}}

\def\psik{|\psi\rangle}
\def\psibr{\langle\psi|}

\def\epsilonk{|\epsilon\rangle}

\def\EE{{\cal E}}
\def\FF{{\cal F}}
\def\LL{{\cal L}}
\def\MM{{\cal M}}
\def\DD{{\cal D}}
\def\VV{{\cal V}}
\def\smD{{\scriptscriptstyle D}}

\def\Dline{{D \kern-0.6em  /  }\,\, }
\def\DDline{{{\cal D} \kern-0.6em  /  }\,\, }
\def\parline{\,\partial\kern -0.55em /\,\,}

\def\fwh{\widehat{f}}

\def\fwt{\widetilde{f}}
\def\mwt{\widetilde{m}}

\def\sma{{\scriptscriptstyle (a)}}
\def\smone{{\scriptscriptstyle (1)}}
\def\smtwo{{\scriptscriptstyle (2)}}
\def\smthree{{\scriptscriptstyle (3)}}
\def\smfour{{\scriptscriptstyle (4)}}

\def\mas{{\rm m}}
\def\coscon{{\rho}}
\def\im{{\rm im}}

\def\LL{{\cal L}}
\def\MM{{\cal M}}

\def\wlin{{\rm w}}

\def\half{{\frac{1}{2}}}

\begin{document}


\begin{flushright}
FIAN/TD/13-2006 \\
hep-th/0612279 \ \ \ \ \ \
\end{flushright}

\vspace{1cm}

\begin{center}

{\Large \bf Gravitational and higher-derivative interactions

\bigskip
of massive spin 5/2 field in  (A)dS space}

\vspace{2.5cm}

R.R. Metsaev\footnote{ E-mail: metsaev@lpi.ru }

\vspace{1cm}

{\it Department of Theoretical Physics, P.N. Lebedev Physical
Institute, \\ Leninsky prospect 53,  Moscow 119991, Russia }

\vspace{3.5cm}

{\bf Abstract}

\end{center}

\noindent

Using on-shell gauge invariant formulation of relativistic dynamics
we study interaction vertices for a massive spin 5/2 Dirac field
propagating in (A)dS space of dimension greater than or equal to
four. Gravitational interaction vertex for the massive spin 5/2 field and
all cubic vertices for the massive spin 5/2 field and massless spin 2
field with two and three derivatives are obtained. In dimension
greater that four, we demonstrate that the gravitational vertex of
the massive spin 5/2 field involves, in addition to the standard minimal
gravitational vertex, contributions with two and three derivatives.
We find that for the massive spin 5/2 and massless spin 2 fields one can
build two higher-derivative vertices with two and three derivatives.
Limits of massless and partial massless spin 5/2 fields in (A)dS space and
limits of massive and massless spin 5/2 fields in flat space are
discussed.

\newpage
\renewcommand{\thefootnote}{\arabic{footnote}}
\setcounter{footnote}{0}

\newsection{Introduction and Summary}

\subsection{Motivation for theory of interacting
massive higher-spin fields in (A)dS}

Higher-spin field theories \cite{Fronsdal:1978vb,Vasiliev:2003ev}
formulated in anti-de Sitter space-time have attracted considerable
interest for a long period of time. Conjectured duality
\cite{Maldacena:1997re} of conformal ${\cal N}=4$ SYM theory and
superstring theory in $AdS_5 \times S^5$ background has triggered intensive
study of higher-spin field dynamics in $AdS$ space. Modern higher-spin
fields research can be divided in two themes at least. One of the
themes deals with {\it massless} higher-spin fields in $AdS$
(for recent review, see \cite{Vasiliev:2004cp},\cite{Bekaert:2005vh}),
while another one deals with {\it massive} higher-spin fields in $AdS$.
It is obvious that studying the massive higher-spin fields in $AdS$ should have
certain interesting and valuable applications in string theory
because one expects that space of states of the short superstring in
$AdS$ space is formed by massless low spin fields and massive higher-spin
fields (see e.g. \cite{Tseytlin:2002gz}). One of the interesting
problems of the massive higher-spin fields in $AdS$ is introduction of the
gravitational interaction. In contrast to completeness of description
of the gravitational interaction for massless higher-spin fields in $AdS$
\cite{Vasiliev:2003ev}, not much is yet known on the gravitational
interaction of the massive higher-spin fields in $AdS$. In this paper, we
make first step in this direction by studying the gravitational
interaction for simplest case of massive higher-spin field which is
spin 5/2 field. Also, we study higher-derivative cubic interaction vertices for
the massive spin 5/2 field and massless spin 2 field with two and three
derivatives and obtain complete list of such vertices.

To summarize, theories of interacting massive higher-spin fields in
$AdS$ background can be interesting for two reasons at least. The
first reason is their possible applications in the theory of superstring in
$AdS_5\times S^5$ Ramond-Ramond background. The second reason is that
the massless limit of these theories might have interesting relations
to the massless higher-spin field theories of Ref.\cite{Vasiliev:2003ev}
which, in turn, according to conjecture in Ref.\cite{hagsun,sun}, have
certain dual description in terms of {\it free} large $N$ conformal
${\cal N}=4$ SYM theory%
\footnote{ Interesting recent discussion of the tensionless limit of
strings in flat and $AdS$ spaces may be found in
\cite{Bonelli:2003kh,Bonelli:2003zu}.}.

Before we proceed to the main theme of this paper, let us mention
briefly the gauge invariant approaches which could be used to discuss
interaction vertices of the spin 5/2 field. Since the works
\cite{Fronsdal:1978vb},\cite{Lopatin:1987hz,Vasiliev:1987tk} devoted
to massless fields in $AdS$ space various descriptions of massive and
massless arbitrary spin fields in $(A)dS$ were developed. In
particular, an  ambient space formulation was discussed in
\cite{Metsaev:1995re}-\cite{Fotopoulos:2006ci} and various BRST
formulations were studied in
\cite{Buchbinder:2001bs}-\cite{Buchbinder:2006nu}. The frame-like
formulations of free fields which seems to be the most suitable for
formulation of the theory of interacting fields in $(A)dS$ was
developed in \cite{Alkalaev:2003qv,Alkalaev:2005kw}. Other
interesting formulations of higher-spin field theories were also discussed
recently in \cite{deMedeiros:2003px}-\cite{Baekler:2006vw}. In this
paper we adopt the approach of
Refs.\cite{Zinoviev:2001dt,Metsaev:2006zy} which turns out to be the
most useful for our purposes.

\subsection{Gravitational interaction vertex and
higher-derivative vertices}\label{gravhighver}

The gravitational interaction  of low spin fields, $s=0,\,1/2,\,1,\,3/2$,
is obtained by standard procedure via
covariantization of their free actions%
\footnote{ Recent discussion of cubic vertices for massive and
massless spin 2 fields in the framework of gauge
invariant approach may be found in \cite{Zinoviev:2006im}. Complete
classification of cubic vertices for all massless and massive
arbitrary spin symmetric fields in flat space
was obtained in the framework of light-cone
approach in \cite{Metsaev:2005ar}.}.
By now it is well known \cite{Aragone:1979hx,Berends:1979kg} that the
`naive' gravitational interaction of higher-spin fields, $s>2$,
introduced by using procedure of the covariantization of their free
actions in flat space, turns out to be inconsistent. The reason is
that, in curved background, the gauge variation of the higher-spin field
action contains the Weyl tensor which cannot be compensated by variation
of the graviton metric tensor. It turns out \cite{Fradkin:1987ks}
that in order to get consistent gravitational interaction
of higher-spin fields it is
necessary to add suitable higher-derivative contributions to the minimal
gravitational interaction.

Throughout this paper we refer to {\it Lagrangian obtained by using
procedure of the covariantization of the free field
Lagrangian in flat space
and taken at cubic order in fields
as minimal gravitational vertex}.
In this paper we deal with cubic vertices that
involve two fermionic fields and one massless spin 2 field.
As we have said, the minimal gravitational vertex of
higher-spin field, $s>2$,  breaks higher-spin gauge symmetries and
therefore is not consistent. {\it The minimal gravitational vertex
supplemented by higher-derivative contributions required for the
consistency of the higher-spin gauge symmetries is referred to as
gravitational vertex} in this paper. In general, besides the
gravitational vertex, there are {\it vertices, to cubic order in
fields, that {\bf i}) involve higher derivatives; {\bf ii})
do not involve the minimal
gravitational vertex; {\bf iii}) respect the linearized higher-spin gauge
symmetries}. We refer to such vertices as {\it higher-derivative
vertices}.

In this paper, we build the gravitational vertex for the massive Dirac spin
5/2 field and higher-derivative vertices for the massive Dirac spin
5/2 field and the massless spin 2 field propagating in $(A)dS_d$ space.
These vertices constructed out two massive spin 5/2 fields, one
massless spin 2 field, and covariant derivatives acting on the
fields. In general, a cubic vertex involves contributions with
different powers of derivatives. Therefore, to classify cubic vertices
we need two labels at least. We use labels $k_{min}$ and $k_{max}$
which are the respective minimal and maximal numbers of derivatives
appearing in the cubic vertex.
In general, $k_{min}$ and $k_{max}$ depend
on a dimension of space-time.

In this paper we deal with vertices subject to the following conditions: {\bf i})
vertices should be parity invariant%
\footnote{ We refer to vertex that does not involve one antisymmetric
Levy-Civita symbol and the matrix $\gamma^{d+1}=\gamma^0\ldots \gamma^{d-1}$
as parity invariant vertex.};
{\bf ii}) vertices should have $k_{max}\leq 3$; {\bf iii})
two-derivative and three-derivative
vertices should depend on the massless spin 2 field through the
linearized Weyl tensor of the massless spin 2 field; {\bf iv}) vertices should be
nontrivial on-shell.
We begin with discussion of our results for vertices in dimension greater than
four.

In dimensions $d>4$, we find that:

\medskip
\noindent
{\bf a}) Gravitational vertex for the massive spin 5/2 Dirac field in $(A)dS_d$
(or flat space) involves, in addition to the standard minimal
gravitational vertex, contributions with two and three derivatives,
i.e., the gravitational vertex has $k_{min}=1$, $k_{max}=3$.
The gravitational vertex for the massive spin 5/2 field in $(A)dS_d$
has smooth flat space limit.

\noindent
{\bf b}) For massive spin 5/2 Dirac field and massless spin 2 field in
$(A)dS_d$ (or flat space), one can build two higher-derivative
vertices with two and three derivatives. Both these vertices have
$k_{min}=2$, $k_{max}=3$.

\noindent
{\bf c}) Gravitational vertices for the massless and partial massless spin 5/2
Dirac fields in $(A)dS_d$ involve, in addition to the
minimal gravitational vertices, contributions with two and three
derivatives%
\footnote{ Interesting discussion of the gravitational vertex of
massless spin 5/2 field in $AdS$ may be found in
\cite{Sorokin:2004ie}.},
i.e., for these gravitational vertices we obtain
$k_{min}=1$, $k_{max}=3$.

\noindent
{\bf d}) For massless spin 5/2 Dirac field and the massless spin 2 field in
$(A)dS_d$, there are no higher-derivative vertices with two and three
derivatives.

\noindent
{\bf e}) For partial massless spin 5/2 Dirac field and the massless spin 2
field in $(A)dS_d$, one can build
one higher-derivative vertex subject to
the condition $k_{max}\leq 3$.
The vertex has $k_{min}=2$, $k_{max}=3$.

\noindent
{\bf f}) For massless spin 5/2 Dirac field and the massless spin 2
field in flat space, one can build one higher-derivative vertex subject to
the condition $k_{max}\leq 3$. The vertex has $k_{min}=k_{max}=3$.

\bigskip

\noindent{\sf Table I. Gravitational and higher-derivative parity
invariant vertices%
\footnote{ In Table I, the no-go theorem for
gravitational interaction of massless spin 5/2 field
in flat space was obtained
in Refs. \cite{Aragone:1979hx,Berends:1979kg}. Gravitational
vertex for the massive spin 5/2 field in flat $4d$
was found in \cite{Porrati:1993in}.
}
for Dirac spin 5/2 field in $(A)dS$ and flat spaces
of dimension $d\geq 4$.
$k_{min}$ and $k_{max}$ denote the respective minimal and maximal
numbers of derivatives in the
vertices. In the Table, we present the higher-derivative vertices that
depend on the massless spin 2 field through the linearized Weyl tensor
of the massless spin 2 field and satisfy the condition $k_{max} \leq 3$ .}

{\small
\begin{center}
\begin{tabular}{|c|c|c|c|c|}
\hline &&&& \\[-3mm]
Type of  & Geometry  & $(k_{min}, k_{max})$  &  Number of  &
$(k_{min}, k_{max})$ for
\\
spin 5/2   & and dimension & for gravitational & higher-deriv. &
higher-derivative
\\
Dirac field & of space-time & vertex & vertices & vertices
\\ [1mm]\hline
&&&&
\\[-3mm]
massive & (A)dS, $d>4$ & (1,3)  &   2  & (2,3)
\\[0mm]
&&&& for both vertices
\\[2mm]\hline
&&&&
\\[-3mm]
massive & (A)dS, $d=4$ & (1,2)  &   1  & (2,3)
\\[2mm]\hline
&&&&
\\[-3mm]
massless   & (A)dS, $d>4$ & (1,3)  & -  & -
\\[2mm]\hline
&&&&
\\[-3mm]
massless   & (A)dS, $d=4$ & (1,2)  & - & -
\\[2mm]\hline
&&&&
\\[-3mm]
partial   & (A)dS, $d>4$ & (1,3)  & 1 & (2,3)
\\[-1mm]
massless &  &  &  &
\\[2mm]\hline
&&&&
\\[-3mm]
partial   & (A)dS, $d=4$ & (1,2)  & - &  -
\\[-1mm]
massless &  &  &  &
\\[2mm]\hline
&&&&
\\[-3mm]
massive & Flat, $d>4$ & (1,3)  &   2  & (2,3)
\\[0mm]
&&&& for both vertices
\\ \hline
&&&&
\\[-3mm]
massive & Flat, $d=4$ & (1,2)  &   1  & (2,3)
\\[2mm]\hline
&&&&
\\[-3mm]
massless & Flat, $d>4$ & -  &   1  & (3,3)
\\[2mm]\hline
&&&&
\\[-3mm]
massless & Flat, $d=4$ & -  &   1  & (2,2)
\\[2mm]\hline
\end{tabular}
\end{center}
}

\medskip

In dimension $d=4$, we find that:
\\
{\bf a}) Gravitational vertices for massive, massless, and partial massless
spin 5/2 Dirac fields in $(A)dS_4$  involve, in addition to the
minimal gravitational vertices, higher-derivative
contributions only with two derivatives, i.e., when $d=4$, we obtain
$k_{min}=1$, $k_{max}=2$ for the gravitational vertices%
\footnote{ For the case of massive spin 5/2 field in {\it flat}
space, the same result was obtained in \cite{Porrati:1993in} (see also
\cite{Cucchieri:1994tx}). Discussion of the gravitational interaction
for massive spin 7/2 field in flat space in the framework of string theory
may be found in \cite{Giannakis:1998wi} (see also
\cite{Argyres:1989cu}-\cite{Klishevich:1998sr} for studying string
theory interaction vertices for various fields). Arbitrary integer
spin fields in homogeneous electromagnetic field and in symmetrical
Einstein space were studied in
\cite{Klishevich:1998ng},\cite{Klishevich:1998wr} (see also
\cite{Klishevich:1997pd}-\cite{Buchbinder:1999ar} for related
studies).}.
The gravitational vertex for the massive spin 5/2 field in $(A)dS_4$
has smooth flat space limit.

\noindent
{\bf b}) For massive spin 5/2 Dirac field and
the massless spin 2 field in $(A)dS_4$
(or flat space), one can
build only one higher-derivative vertex with two and three
derivatives, i.e., the vertex has $k_{min}=2$, $k_{max}=3$.

\noindent
{\bf c}) For massless spin 5/2 Dirac field and the massless spin 2 field in
$(A)dS_4$, there are no higher-derivative vertices with two and three
derivatives, i.e., there are no higher-derivative vertices subject
to the condition $ 2\leq k_{min}\leq k_{max}\leq 3$.

\noindent
{\bf d}) For partial massless spin 5/2 Dirac field and the massless spin 2
field in $(A)dS_4$, there are no higher-derivative vertices with two
and three derivatives, i.e., there are no higher-derivative vertices subject
to the condition $ 2\leq k_{min}\leq k_{max}\leq 3$.

\noindent
{\bf e}) For massless spin 5/2 Dirac field and the massless spin 2
field in flat space, there is only one higher-derivative
vertex subject to the condition
$k_{max}\leq 3$. The vertex has $ k_{min} = k_{max} =2$.

\medskip
The vertices above listed
are discussed in Sections \ref{subsubgravvermass01nn}--\ref{secmasverd401}
and summarized in Table I. As byproduct, in Section \ref{speintver01}, we find
some higher-derivative vertices describing
interactions of spin 5/2 field
with low spin, 3/2 and 1/2, fields and the massless spin 2 field.
The latter vertices are summarized in Table II.

\bigskip

\noindent{\sf Table II. Higher-derivative parity
invariant vertices for Dirac spin 5/2, 3/2, 1/2 fields and
massless spin 2 field in $(A)dS$ and flat spaces
of dimension $d\geq 4$.
Subscripts $0$, $m$, and $pm$ denote the
respective massless, massive and partial massless fields.
In this Table, vertices take $k_{min}=k_{max}$.
}%

{\small
\begin{center}
\begin{tabular}{|c|c|c|c|}
\hline &&& \\[-3mm]
Spin values & Geometry  &   Number of  &
number
\\
of fields    & and dimension &  higher-deriv. &
of derivatives
\\
in cubic vertex & of space-time &  vertices & in vertices
\\ [1mm]\hline
&&&
\\[-3mm]
$\left(\frac{5}{2}\right)_0-\left(\frac{3}{2}\right)_0-2_0$
& flat, $d>4$ &    2  & 3 \ (for both vertices)
\\
& flat, $d=4$ &    -  & -
\\[2mm]\hline
&&&
\\[-3mm]
$\left(\frac{5}{2}\right)_0-\left(\frac{1}{2}\right)_0-2_0$
& flat, $d\geq 4$ & 2  & 3 \ (for both vertices)
\\[2mm]\hline
&&&
\\[-3mm]
$\left(\frac{5}{2}\right)_{pm}-\left(\frac{1}{2}\right)_m-2_0$
& (A)dS, $d\geq4$ & 2 & 3 \ (for both vertices)
\\[-2mm]
  &  &  &
\\[2mm]\hline
\end{tabular}
\end{center}
}

\medskip

We now present our result for the cubic interaction vertices. We
discuss the interaction vertices in the framework of on-shell gauge
invariant approach. In on-shell gauge invariant approach, the massive
spin 5/2 Dirac field is described by Dirac complex-valued
tensor-spinor fields $\psi^{AB\alpha}$, $\psi^{A\alpha}$
$\psi^\alpha$%
\footnote{ $A,B=0,1,\ldots,
d-1$ are flat vector indices of the $so(d-1,1)$ algebra.
Fields with flat indices $\psi^{AB}$ are defined in terms
of base manifold tensor fields $\psi^{\mu\nu}$ as $\psi^{AB} \equiv
e_\mu^A e_\nu^B \psi^{\mu\nu}$, where
$e_\mu^A$ is the vielbein of $(A)dS$ background.}.
The fields  $\psi^{AB\alpha}$, $\psi^{A\alpha}$,
$\psi^\alpha$ transform in the respective spin $5/2$, $3/2$, $1/2$
non-chiral representations
of the $so(d-1,1)$ Lorentz algebra and satisfy appropriate differential
and algebraic
constraints (for details, see Section \ref{ongauformsec01}).
In what follows, we often suppress the spinor index $\alpha$.
We shall refer to the field $\psi^{AB}$ as generic field, while the fields
$\psi^{A}$ and $\psi$ will be referred to as Stueckelberg fields. We use
oscillators $\alpha^A$, $\zeta$
to collect the fields in a ket-vector $\psik$ defined by
\be\label{psikexp01}
\psik  \equiv \left( \alpha^A\alpha^B \psi^{AB}  + \zeta \alpha^A
\psi^A
  +  \zeta^2 \psi \right)|0\rangle\,.
\ee

To cubic order in fields (two massive spin 5/2 fields and one massless spin
2 field), Lagrangian  involving one, two, and three derivatives can
be written as
\be\label{Lint}
\LL^{int} = \LL_\smone + \LL_\smtwo + \LL_\smthree\,,
\ee
where $\LL_\smone$ stands for the minimal gravitational vertex, while
$\LL_\smtwo$ and $\LL_\smthree$ stand for the respective vertices with
two and three derivatives. All these vertices can be represented as%
\footnote{The bra-vector $\psibr$ is defined according the rule
$\psibr = (\psik)^\dagger\gamma^0$. We use $e\equiv \det e_\mu^A$.}
\be \label{LLaMMadef01}
{\rm i}e^{-1} \LL_\sma = \psibr M_\sma \psik\,, \qquad\quad a=1,2,3\,,
\ee
where
operators $M_\sma$ are constructed out the on-shell $(A)dS$
massless spin 2 field, denoted by $h^{AB}$, covariant derivatives,
the oscillators $\alpha^A$, $\bar\alpha^A$, $\zeta$, $\bar\zeta$, and the
Dirac $\gamma$-matrices. We refer to these operators as vertex operators.
The vertex operator $M_\smone$ describes {\it the
minimal gravitational interaction} of the massive spin 5/2 field and it can be
obtained by using the standard procedure of covariantization of the
massive spin 5/2 field free action in flat space and expanding over
$(A)dS$ background%
\footnote{Because the ket-vector $\psik$ is assumed to be $\gamma$-traceless,
we drop $\gamma^A\alpha^A$, and $\gamma^A\bar\alpha^A$ terms
in $M_\smone$,\rf{mingravverdef01}.},
\be \label{mingravverdef01}
M_\smone = -\frac{1}{2}h^{AB} \gamma^A D^B +
\wlin^{ABC}\gamma^A\alpha^B\bar\alpha^C\,,
\ee
where $D^A$ is covariant derivative acting on the ket-vector $\psik$
(see \rf{lorspiope}) and $\wlin^{ABC}=-\wlin^{ACB}$ is
the linearized Lorentz connection constructed out the on-shell massless
spin 2 field,
\be \label{omegalin01}
\wlin^{ABC}= \frac{1}{2}(- \DD^B h^{AC}  +  \DD^C h^{AB})\,.
\ee
In \rf{omegalin01} and below, $\DD^A$  stands for the covariant
derivative acting on fields with flat indices (see Appendix A).
The vertex operators
$M_\smtwo$ and $M_\smthree$ involving the respective two and three
derivatives can be expanded in terms of base vertex operators
\beq\label{Mtwo}
&& M_\smtwo = \sum_{a=1}^3 M_{2,a} + M_{2,3}^\im\,,
\\
\label{Mthree} && M_\smthree = \sum_{a=1}^5 M_{3,a} + \sum_{a=3}^5
M_{3,a}^\im\,,
\eeq
where the respective two-derivative and three-derivative base vertex operators
$M_{2,a}$, $M_{2,3}^\im$ and $M_{3,a}$, $M_{3,a}^\im$ take the form%
\footnote{ Throughout this paper symmetrization of indices $(AB)$ is normalized as $(AB)=
\frac{1}{2}(AB + BA)$.}:
\beq
\label{M21def} && M_{2,1} = m_{2,1} C^{ABCE} \gamma^{AB} \alpha^C
\bar\alpha^E \,,
\\[3pt]
&& M_{2,2} = m_{2,2} C^{ABCE} \alpha^A\alpha^E
\bar\alpha^B\bar\alpha^C \,,
\\[3pt]
&& M_{2,3} = m_{2,3} C^{ABCE} \gamma^A \left( \alpha^B \alpha^C
\bar\alpha^E \bar\zeta
- \zeta \alpha^E \bar\alpha^B \bar\alpha^C \right)\,,
\\[15pt]
\label{romM31def01}
&& M_{3,1} = m_{3,1} C^{A(BC)E} \gamma^A \alpha^B \bar\alpha^C D^E\,,
\\[3pt]
&& M_{3,2} = \frac{1}{2}m_{3,2} \DD^F C^{ABCE} \gamma^A (\alpha^B
\alpha^C \bar\alpha^E\bar\alpha^F - \alpha^E\alpha^F\bar\alpha^B
\bar\alpha^C)\,,
\\[3pt]
&& M_{3,3} = C^{ABCE} \gamma^{AB} \left(  m_{3,3} \alpha^C \bar\zeta
+  \zeta \bar\alpha^C  m_{3,3}\right) D^E\,,
\\[3pt]
&& M_{3,4} = m_{3,4} C^{ABCE} \left( \alpha^B\alpha^C \bar\alpha^A
\bar\zeta
- \zeta \alpha^A \bar\alpha^B \bar\alpha^C \right) D^E\,,
\\[3pt]
&& M_{3,5} = m_{3,5} C^{ABCE} \gamma^A \left( \alpha^B \alpha^C
\bar\zeta^2
+ \zeta^2 \bar\alpha^B \bar\alpha^C \right) D^E\,,
\eeq
\beq
&& M_{2,3}^\im = m_{2,3}^\im C^{ABCE} \gamma^A \left( \alpha^B \alpha^C
\bar\alpha^E \bar\zeta
+ \zeta \alpha^E \bar\alpha^B \bar\alpha^C \right)\,,
\\[3pt]
&& M_{3,3}^\im = C^{ABCE} \gamma^{AB} \left(  m_{3,3}^\im \alpha^C
\bar\zeta
-  \zeta \bar\alpha^C  m_{3,3}^\im\right) D^E\,,
\\[3pt]
&& M_{3,4}^\im = m_{3,4}^\im C^{ABCE} \left( \alpha^B\alpha^C
\bar\alpha^A \bar\zeta
+ \zeta \alpha^A \bar\alpha^B \bar\alpha^C \right) D^E\,,
\\[3pt]
\label{M35cdef} && M_{3,5}^\im = m_{3,5}^\im C^{ABCE} \gamma^A \left(
\alpha^B \alpha^C \bar\zeta^2
- \zeta^2 \bar\alpha^B \bar\alpha^C \right) D^E\,.
\eeq
Here and below, $C^{ABCE}$ stands for the linearized Weyl tensor constructed out
the on-shell massless spin 2 field $h^{AB}$ and two derivatives.
Quantities $m_{a,b}$, $m_{a,b}^\im$ are independent of the oscillators
$\alpha^A$, $\bar\alpha^A$ and the $\gamma$-matrices. Some of these
quantities depend on the
operator $N_\zeta\equiv \zeta\bar\zeta$. This,
\be m_{2,2}, \ m_{2,3},\ m_{3,2},\ m_{3,4}, \ m_{3,5},\ m_{2,3}^\im,
m_{3,4}^\im,\ m_{3,5}^\im \ \ \ \hbox{ do not depend on } N_\zeta\,, \ee
while the remaining $m_{a,b}$, $m_{a,b}^\im$ take the form
\beq \label{mabexp01}
&& m_{2,1} = m_{2,1}(0)(1-N_\zeta) + m_{2,1}(1) N_\zeta\,,
\\[3pt]
&& m_{3,1} = m_{3,1}(0)(1-N_\zeta) + m_{3,1}(1) N_\zeta\,,
\\[3pt]
&& m_{3,3} = m_{3,3}(0)(1-N_\zeta) + m_{3,3}(1) N_\zeta\,,
\\[3pt]
\label{mabexp04} && m_{3,3}^\im = m_{3,3}^\im(0)(1-N_\zeta) +
m_{3,3}^\im(1) N_\zeta\,.
\eeq
Expansions \rf{mabexp01}-\rf{mabexp04} are defined so that the coefficients
$m_{a,b}(0)$ and $m_{a,b}(1)$ turn out to the respective
values of $m_{a,b}$ for $N_\zeta=0$ and $N_\zeta=1$. We note
that
\beq \label{mabreacon01}
&&
m_{a,b}, \ \ m_{a,b}(0), \ \ m_{a,b}(1) \qquad \hbox{ are real valued numbers}\,;
\\
&& \label{mabreacon02}
m_{a,b}^\im, \ \ m_{a,b}^\im(0), \ \ m_{a,b}^\im(1) \qquad
\hbox{ are pure imaginary numbers.}
\eeq

{\it The gravitational vertex} of the massive spin 5/2 field is described
by the operator $M_\smone$ supplemented by the operators
$M_{2,a}$, $M_{3,a}$ with appropriate choice of
$m_{a,b}$. Higher-derivative vertices with two and three
derivatives are described by the respective operators $M_{2,a}$, $M_{2,3}^\im$ and
$M_{3,a}$, $M_{3,a}^\im$%
\footnote{ As an side of remark, we note that vertices
related to the operators $M_{2,3}^\im$, $M_{3,a}^\im$ vanish when the
Dirac spin 5/2 field $\psik$ is restricted to the Majorana spin 5/2
field.}.
Note that the operators $M_{2,a}^\im$, $M_{3,a}^\im$ do not contribute to
the gravitational vertex.

Making use of simplified notation for the generic field $\psi^{AB}$ and
Stueckelberg fields $\psi^A$, $\psi$, \rf{psikexp01},
\be \label{psisimnot01} \psi_2 \sim \psi^{AB}\,, \qquad \psi_1 \sim \psi^A \,,
\qquad
\psi_0 \sim \psi\,, \ee
and dropping $\gamma$-matrices and tensor indices, the vertices
can schematically be represented as
\beq
\LL_\smone & \sim  & \psi_2 (hD +\wlin) \psi_2 + \psi_1 (hD+\wlin)
\psi_1 + \psi_0 h D \psi_0\,,
\\[7pt]
\label{LLtwodef01}
\LL_\smtwo & \sim  & m_{2,1}(0) \psi_2 C \psi_2 + m_{2,2}\psi_2 C \psi_2
+ m_{2,1}(1) \psi_1 C
\psi_1 + m_{2,3}\psi_2 C \psi_1+ m_{2,3}^\im\psi_2 C \psi_1\,,\qquad
\\[7pt]
\label{LLthreedef01}
\LL_\smthree & \sim &  m_{3,1}(0) \psi_2 C D \psi_2  + m_{3,2} \psi_2
\DD C \psi_2
\nonumber\\
& + & m_{3,1}(1) \psi_1 C D \psi_1 + m_{3,3}(0) \psi_2 C D \psi_1 +
m_{3,4}\psi_2 C D \psi_1
\nonumber\\
& + &  m_{3,3}(1) \psi_1 CD\psi_0 + m_{3,5}\psi_2 C D\psi_0
\nonumber\\
& + &  m_{3,3}^\im(0) \psi_2 CD\psi_1 + m_{3,4}^\im\psi_2 C D\psi_1
%
%
 +  m_{3,3}^\im(1) \psi_1 CD\psi_0 + m_{3,5}^\im\psi_2 C D\psi_0\,,\qquad
\eeq
where $C$ and $D$ stand for the Weyl tensor and the covariant
derivative respectively. This, all that is
required for finding the Lagrangian is to determine
the quantities $m_{a,b}$ and $m_{a,b}^\im$.

Our approach allows us to study vertices for fields in $(A)dS_d$ and
flat spaces on equal footing. In $(A)dS_d$ space, there are two
limits for the massive spin 5/2 field: the limit of massless field and
the limit of partial massless field%
\footnote{Recent studying the partial massless fields
\cite{Deser:1983mm} may be found in
\cite{Deser:2001xr,Deser:2001us,Skvortsov:2006at,Metsaev:2006zy}.}.
To discuss these limits, we introduce
quantities $\fwt_1(0)$, $\fwt_1(1)$ defined by%
\footnote{ The quantities $\fwt_1(0)$, $\fwt_1(1)$ arise in the gauge
transformation of massive spin 5/2 field \cite{Metsaev:2006zy} (see
also Section \ref{ongauformsec01}). Their zero values
determine notion of masslessness and partial masslessness in $(A)dS$ space
(see \rf{caseads2},\rf{caseads3}).}
\beq \label{fwtonezerdef}
&& \fwt_1(0) = \left( \frac{d+1}{d} \Bigl(\mas^2 + \frac{d^2}{4}
\coscon \Bigr)\right)^{1/2}\,,
\\
&& \fwt_1(1) = \left( \frac{d}{d-2} \Bigl(\mas^2 + \frac{(d-2)^2}{4}
\coscon \Bigr)\right)^{1/2}\,,
\eeq
where $\mas$ is a mass of the spin 5/2 field, while $\coscon$ is
expressible in terms of a radius of $(A)dS_d$ space $R$ as
\be \label{omegadef}
\coscon = \left\{\begin{array}{cl}
-\frac{1}{R^2} & \hbox{for AdS space},
\\[5pt]
0 & \hbox{for flat space},
\\[5pt]
+ \frac{1}{R^2} & \hbox{for dS space}.
\end{array}\right.
\ee

\bigskip

Depending on the values of mass parameter $\mas$ and the parameter $\coscon$,
there are the following possibilities%
\footnote{In addition to the possibilities in \rf{caseads1}-\rf{caseads3},
there is some special massive
field in $(A)dS$ corresponding to $\fwt_1(0) \ne 0$, $\fwt_1(1) \ne 0$,
$\mas=0$. In $AdS$ space, this special massive field is not
related to unitary representations of $AdS$ space-time symmetry algebra.
Therefore, to keep discussion from becoming unwieldy,
here we do not consider this special massive field.}:

\bigskip

{\bf Spin 5/2 field in (A)dS space, $\coscon \ne 0$,}
\beq \label{caseads1}
&&\fwt_1(0) \ne 0\,, \quad \fwt_1(1) \ne 0\,, \quad \mas \ne
0\,,\hspace{2.2cm} \hbox{ massive field in (A)dS};
\\[5pt]
\label{caseads2}
&& \fwt_1(0) = 0\,, \quad \fwt_1(1) \ne 0\,, \quad \mas =
\frac{d}{2}\sqrt{- \coscon}\,, \hspace{1.2cm} \hbox{ massless field
in (A)dS};
\\
\label{caseads3} && \fwt_1(0) \ne 0\,, \quad \fwt_1(1) = 0\,, \quad
\mas  = \frac{d-2}{2} \sqrt{- \coscon} \,,
\hspace{0.5cm}  \hbox{ partial massless field in (A)dS}\,. \hspace{1cm}
\
\eeq

\bigskip

{\bf Spin 5/2 field in flat space, $\coscon = 0$,}
\beq
&& \mas \ne 0\,,\hspace{2cm} \hbox{massive field in flat space};
\\
&& \mas = 0\,, \hspace{2cm}  \hbox{massless field in flat space}\,.
\eeq

As we have said, the Lagrangian is completely determined by
the coefficients $m_{a,b}$, $m_{a,b}^\im$. We find these coefficients by
requiring the vertices to be gauge invariant.
From now on, we discuss vertices for the
massive, massless and partial massless field in turn.
In some contexts, there are some differences between vertices in $d>4$ and
the ones in $d=4$. Therefore we separately consider the cases $d>4$
and $d=4$.

The rest of the paper is organized as follows.
In Sections \ref{subsubgravvermass01nn}--\ref{secuniver01},
we discuss vertices for the massive, massless and partial
massless fields in $d>4$.
Section \ref{secmasverd401} is devoted to vertices
for the massive, massless and partial massless fields in $d=4$.
Results in Sections \ref{subsubgravvermass01nn}--\ref{secmasverd401}
are summarized
in Table I. In Section \ref{speintver01}, we discuss
cubic vertices describing interactions of the spin 5/2 field
with low spin 3/2 and 1/2 fields and the massless spin 2 field.
Results in Section \ref{speintver01} are summarized
in Table II. In Section \ref{ongauformsec01}
we discuss the
on-shell gauge invariant formulation for the massive spin
5/2 field and the massless spin 2 field we use in this paper.
In Section \ref{ResSec01}, we analyze
restrictions imposed on the gravitational and higher-derivative vertices
by on-shell gauge symmetries. In Appendix A, we present our notation and
conventions. Helpful commutators between
various quantities constructed out the covariant derivative and the oscillators
are collected in Appendix B.
In Appendix C, we outline
on-shell gauge invariant formulation for massive arbitrary spin
fermionic field and present some details of derivation of
on-shell gauge invariant formulation for the massive spin 5/2
field.  Derivation of gauge variation of
the interaction vertices is discussed in Appendix D.
In Appendices E,F, we discuss bases of the vertex operators
for $d\geq4$ and some special relations appearing
for the interaction vertices when $d=4$.

\newsection{Massive spin 5/2 field, $d>4$}\label{subsubgravvermass01nn}

\subsection{Gravitational vertex for massive spin 5/2 field in $(A)dS_d$,
$d>4$}\label{subsubgravvermass01}

We begin with discussion of the gravitational
vertex for the massive spin 5/2 field in $AdS_d$, \rf{caseads1}.
For this case, solution to the equations%
\footnote{ These equations are given in
\rf{m21repnn}-\rf{m3eq3} and
\rf{m3eq1cnn}-\rf{m3eq5c}.}
for
$m_{a,b}$, $m_{a,b}^\im$
required by on-shell gauge invariance can uniquely be determined
by imposing the conditions $m_{2,2}=0$, $m_{3,2}=0$.
In this way, we obtain
\beq
\label{gravmassiv01} && m_{2,1}(0) = 0\,,
\\[3pt]
&& m_{2,1}(1) = \frac{3\mas}{2d\fwt_1^2(0)}\,,
\\[3pt]
&& m_{2,2} = 0\,,
\\[3pt]
&& m_{2,3} = -\frac{1}{\fwt_1(0)} \,,
\\
&& m_{3,1}(0) = 0\,,
\\
&& m_{3,1}(1) = \frac{1}{\fwt_1^2(0)}\,,
\\[3pt]
&& m_{3,2} =  0 \,,
\\[3pt]
&& m_{3,3}(0) =0\,,
\\[3pt]
&& m_{3,3}(1) = \frac{(1-d)\mas}{d(d-2)\fwt_1^2(0)\fwt_1(1)}\,,
\\
&& m_{3,4} = 0\,,
\\
\label{gravmassiv11} && m_{3,5} = \frac{1}{2\fwt_1(0)\fwt_1(1)} \,,
\\[3pt]
\label{gravmassiv12} && m_{a,b}^\im = 0 \,,\qquad \hbox{ for all } \ a,b\,.
\eeq
Adding the minimal gravitational vertex $\LL_\smone$ to
the sum $\LL_\smtwo+\LL_\smthree$ with above-given $m_{ab}$, $m_{a,b}^\im$
gives the gravitational vertex for the massive spin 5/2 field.

The following remarks are in order.

\noindent
{\bf i)}
To restore dependence on the gravitational constant
$\kappa$, we should simply to multiply the r.h.s. of relations
\rf{mingravverdef01} and
\rf{gravmassiv01}-\rf{gravmassiv12} by the factor $\sqrt{2}\kappa$.

\noindent
{\bf ii)} From \rf{gravmassiv01}-\rf{gravmassiv12}, we see that
$m_{2,a}\ne0$, $m_{3,a}\ne0$ for some $a$. This implies that
the gravitational vertex
involves, in addition to the minimal gravitational vertex,
contributions with two and three derivatives, i.e., the gravitational vertex has
$k_{min}=1$, $k_{max}=3$. There is no possibility for
gravitational vertex with $k_{max}=2$, when $d>4$.

\medskip
\noindent
{\bf iii)}
As will be explained below, there are two
higher-derivative vertices having
$k_{min}=2$ and $k_{max}=3$. Due to this,
there is two parametric freedom in
choosing the gravitational vertex, i.e., the restrictions
imposed by gauge invariance do not allow us
to determine the gravitational vertex uniquely.
Therefore we need some additional restrictions
which allow us to determine the vertex uniquely.
We impose the
following additional restrictions: $m_{2,2}=0$, $m_{3,2}=0$%
\footnote{ We note that the conditions $m_{2,2}=0$ and $m_{3,2}=0$ amount
to the respective conditions $m_{2,1}(0)=0$ and $m_{3,1}(0)=0$.}.
These
additional restrictions amount to
requiring that $\psi_2 C \psi_2$- and $\psi_2 C D \psi_2$-terms
(see \rf{LLtwodef01},\rf{LLthreedef01})
do not contribute to the gravitational vertex. We note that,
in analysis of the gravitational vertex taken to cubic order in fields,
the particular
choice of $m_{2,2}$, $m_{3,2}$ amounts just to choice of scheme for
the gravitational vertex. Physical values
of $m_{2,2}$, $m_{3,2}$ can be determined by study of complete,
i.e., to all order in fields, theory of massive higher-spin fields.

\medskip
\noindent
{\bf iv)} In the scheme we chose, the
gravitational vertex does not have smooth massless limit,
$\fwt_1(0)\rightarrow 0 $, and
partial massless limit, $\fwt_1(1)\rightarrow 0 $. It turns out however that
$1/(\fwt_1(0)\rightarrow 0) $ singularities of the massless limit and
$1/(\fwt_1(1)\rightarrow 0)$ singularities of the partial massless limit can
be canceled,  as will be demonstrated below,
by appropriate choice of scheme, i.e., by adding
to the gravitational vertex some
combination of the two higher-derivative vertices.

\subsection{ Higher-derivative vertices for massive spin 5/2 field and
massless spin 2 field in $(A)dS_d$, $d>4$}\label{hihdersubsubsec}

As we have said, when $2 \leq k_{min}\leq k_{max} \leq 3$,
one can build only two higher-derivative vertices for the massive
spin 5/2 field and the massless spin 2 field.
We consider these vertices in turn.

{\bf (A) First higher-derivative vertex}. First
solution to the equations for $m_{a,b}$, $m_{a,b}^\im$ required by on-shell
gauge invariance can uniquely be determined by imposing the condition $m_{3,2}=0$
and by equating the gravitational constant
$\kappa$ to zero. In this way, we obtain
\beq
\label{highdera01} && m_{2,1}(0) = \frac{1}{\mas} g \,,
\\[3pt]
&& m_{2,1}(1) = \frac{d-1}{\mas\fwt_1^2(0)}\Bigl(
\frac{d-8}{d^2}\mas^2 + \frac{d-4}{4}\coscon \Bigr) g \,,
\\[3pt]
&& m_{2,2} =  \frac{4}{\mas}g\,,
\\[3pt]
&& m_{2,3} = \frac{4(d-2)}{d\fwt_1(0)} g  \,,
\\[3pt]
&& m_{3,1}(0) = 0\,,
\\[3pt]
&& m_{3,1}(1) = \frac{8(1-d)}{d\fwt_1^2(0)} g \,,
\\[3pt]
&& m_{3,2} =  0 \,,
\\[3pt]
&& m_{3,3}(0) = -\frac{1}{\mas\fwt_1(0)} g \,,
\\[3pt]
&& m_{3,3}(1) = - \frac{1}{\mas\fwt_1^2(0)\fwt_1(1) }\Bigl(
\frac{d^2-10d +12}{d(d-2)}\mas^2 + \frac{d(d-4)}{4} \coscon
\Bigr)g\,,
\\[3pt]
&& m_{3,4} = -\frac{4}{\mas\fwt_1(0)} g\,,
\\[3pt]
\label{highdera11} && m_{3,5} = -\frac{4
(d-3)}{(d-2)\fwt_1(0)\fwt_1(1)} g \,,
\\[3pt]
\label{highdera11n1} && m_{a,b}^\im = 0 \,,\qquad \hbox{ for all } \ a,b\,,
\eeq
where we introduce a coupling constant $g$. Here and below, we normalize
all coupling constants of higher-derivative vertices to dimension
$[mass]^{d-4}$ when possible.

{\bf (B) Second higher-derivative vertex}. For
this case, solution to the equations for $m_{a,b}$, $m_{a,b}^\im$
required by on-shell
gauge invariance can uniquely be determined by imposing
the condition $m_{2,2}=0$,
and by equating the gravitational constant $\kappa$
to zero. In this way, we obtain
\beq
\label{highderb01} && m_{2,1}(0) = 0 \,,
\\[3pt]
&& m_{2,1}(1) = \frac{1}{d\mas\fwt_1^2(0)}\Bigl(
\frac{3d+3}{d^2}\mas^2 + \frac{13-7d}{4} \coscon \Bigr) g' \,,
\\[3pt]
&& m_{2,2} = 0\,,
\\[3pt]
&& m_{2,3} = \frac{1}{\mas^2\fwt_1(0)} \Bigl( - \frac{d+1}{d^2}\mas^2
+ \frac{5d-5}{4} \coscon \Bigr) g'   \,,
\\[3pt]
&& m_{3,1}(0) = \frac{1}{\mas^2}g'\,,
\\[3pt]
&& m_{3,1}(1) = \frac{1}{\mas^2\fwt_1^2(0)}\Bigl(
\frac{d^2+d+2}{d^2}\mas^2 + \frac{d^2-3d+6}{4} \coscon \Bigr) g' \,,
\\[3pt]
&& m_{3,2} =  \frac{1}{\mas^2} g' \,,
\\[3pt]
&& m_{3,3}(0) = -\frac{1}{2\mas d\fwt_1(0)} g'\,,
\\[3pt]
&& m_{3,3}(1) = - \frac{1}{2\mas d(d-2)\fwt_1^2(0)\fwt_1(1) }\Bigl(
(d+2) \mas^2 + \frac{(d-4)(d^2-6d+4)}{4} \coscon \Bigr) g' \,, \ \ \
\ \ \
\\[3pt]
&& m_{3,4} = -\frac{2}{d\mas\fwt_1(0)} g'\,,
\\[3pt]
\label{highderb11} && m_{3,5} =
\frac{1}{\mas^2\fwt_1(0)\fwt_1(1)}\Bigl( \frac{\mas^2}{d-2} +
\frac{4-d}{4} \coscon \Bigr) g' \,,
\\[3pt]
\label{highderb11n1}
&& m_{a,b}^\im = 0 \,,\qquad \hbox{ for all } \ a,b\,,
\eeq
where $g'$ stands for a coupling constant.

The following remarks are in order.

\noindent
{\bf i)} From \rf{highdera01}-\rf{highderb11n1}, we see that
$m_{2,a}\ne0$, $m_{3,a}\ne0$ for some $a$, i.e.,
both higher-derivative vertices
involve two and three derivatives. This, both vertices have
$k_{min}=2$, $k_{max}=3$.

\noindent
{\bf ii)}
Since both higher-derivative vertices have the same
$k_{min}$ and $k_{max}$ we need rule
which allows us, firstly, to distinguish between these two vertices and,
secondly, to determine these vertices uniquely in due course of
solving the restrictions imposed by gauge invariance.
The rule we are using is that the two higher-derivative vertices can
be distinguished by
$\psi_2 C \psi_2$- and $\psi_2 C D \psi_2$-terms, where
$\psi_2$ is the generic field (see \rf{psisimnot01}). This,
the type $(\bf A)$ vertex is uniquely determined by
requiring that the $\psi_2 C D \psi_2$-terms do not contribute to the Lagrangian
(i.e., by imposing the condition $m_{3,2}=0$,
see \rf{LLthreedef01}), while
the type $(\bf B)$ vertex is uniquely determined by
requiring that the $\psi_2 C \psi_2$-terms do not contribute to the Lagrangian
(i.e., by imposing the condition $m_{2,2}=0$,
see \rf{LLtwodef01}).

\noindent
{\bf iii}) Both
higher-derivative vertices are singular in the limit of massless field,
$\fwt_1(0)\rightarrow 0 $,
and the limit of partial massless field, $\fwt_1(1)\rightarrow 0 $.
It turns out however that
$1/(\fwt_1(1)\rightarrow 0) $ singularities of
the partial massless field limit
can be canceled in some special combination of the two higher-derivative vertices
(see Section \ref{subsechigderverparmas01} below).
This gives one higher-derivative vertex for the partial massless field
in $(A)dS_d$, $d>4$.
The $1/(\fwt_1(0)\rightarrow 0)$ singularities of
the massless field limit can in no way
be canceled%
\footnote{
We can re-scale the coupling constants,
$g \rightarrow \fwt_1^2(0) g$, $g' \rightarrow \fwt_1^2(0) g'$,
to get well defined vertices in the limit
of massless field. But, in doing so, we obtain non-trivial
interaction only for spin 3/2 field and massless spin 2 field.}.
This implies that there are
no higher-derivative vertices
for the massless spin 5/2 field
in  $(A)dS_d$, when $k_{max}\leq 3$ and $d>4$.

\subsection{Gravitational and higher-derivative
vertices for massive spin 5/2 field in flat space,
$d>4$}

From expressions for vertices
given in the previous Sections, we see
that the gravitational and higher-derivative vertices for the massive
spin 5/2 field in $(A)dS$ space allow smooth flat space limit.
In this limit, realized as $R\rightarrow \infty$ (see \rf{omegadef}), we have
\beq \label{fwtonezerdeflim01}
&& \rho \rightarrow  0\,,
\\
\label{fwtonezerdeflim02}
&& \fwt_1(0) \rightarrow \left( \frac{d+1}{d}\right)^{1/2}\mas\,,
\\
\label{fwtonezerdeflim03}
&& \fwt_1(1) \rightarrow \left( \frac{d}{d-2}\right)^{1/2} \mas\,.
\eeq
In other words,
the gravitational and higher-derivative vertices for the massive
spin 5/2 field in $(A)dS$ do not have poles at $\rho=0$.
Therefore, all that is required
to obtain the gravitational and higher-derivative vertices
for the massive spin 5/2 field in flat space is to allow the limiting
values for $\rho$, $\fwt_1(0)$, $\fwt_1(1)$
\rf{fwtonezerdeflim01}-\rf{fwtonezerdeflim03} in the respective
expressions for $(A)dS$ vertices.

\newsection{Massless and partial massless spin 5/2 fields, $d>4$}

\subsection{Gravitational vertex for massless spin 5/2 field in $(A)dS_d$,
$d>4$}

The massless spin 5/2 field is described by generic
field $\psi_2$ \rf{psisimnot01} with the mass parameter given in
\rf{caseads2}.
Solution to the equations%
\footnote{ These equations are obtained from the ones given
in \rf{m21repnn}-\rf{m3eq3} and
\rf{m3eq1cnn}-\rf{m3eq5c} by substitution $\fwt_1(0) =  0 $.}
for $m_{a,b}$, $m_{a,b}^\im$
required by on-shell gauge invariance can uniquely be determined
by imposing the conditions $m_{3,1}(1)=0$, $m_{3,3}(1)=0$.
In this way, we get
\beq
\label{masslmm01} && m_{2,1}(0) = \frac{1}{8(d-1)\sqrt{- \coscon
}}\,,
\\[3pt]
&& m_{2,1}(1) = 0\,,
\\[3pt]
&& m_{2,2} = \frac{1}{2 (d-1) \sqrt{- \coscon } }\,,
\\[3pt]
&& m_{2,3} = 0  \,,
\\[3pt]
&& m_{3,1}(0) = \frac{1}{2  (d-1) \coscon }\,,
\\[3pt]
&& m_{3,1}(1) = 0\,,
\\[3pt]
&& m_{3,2} =  \frac{1}{2  (d-1) \coscon } \,,
\\
&& m_{3,3}(0) =0\,, \qquad
m_{3,3}(1) = 0\,,
\\
\label{masslmm11} && m_{3,4} = 0\,, \qquad \quad \ m_{3,5} = 0 \,,
\\
\label{masslmm12} && m_{a,b}^\im = 0\,, \qquad \hbox{ for all } \  a,b\,.
\eeq
Adding the minimal gravitational vertex $\LL_\smone$
to the sum $\LL_\smtwo+\LL_\smthree$ with above-given $m_{ab}$, $m_{a,b}^\im$
and ignoring contribution of the fields $\psi_1$, $\psi_0$ to $\LL_\smone$
gives the gravitational vertex for the massless spin 5/2 field.

The following remarks are in order.

\noindent
{\bf i)} From \rf{masslmm01}-\rf{masslmm12}, we see that
$m_{2,a}\ne0$, $m_{3,a}\ne0$ for some $a$. This implies that
the gravitational vertex
involves, in addition to the minimal gravitational vertex,
contributions with two and three derivatives, i.e., the gravitational
vertex has $k_{min}=1$, $k_{max}=3$.
There is no possibility for gravitational vertex
with $k_{max}=2$ when $d>4$.

\noindent
{\bf ii)} From
\rf{masslmm01}-\rf{masslmm12}, we see that the
gravitational vertex for the massless spin 5/2 field in $(A)dS_d$
does not have smooth
flat space limit, $ \coscon \rightarrow 0$,
\cite{Fradkin:1987ks}. Below, we demonstrate that, in $d=4$, the
gravitational vertex does not involve three-derivative
contributions, i.e., $1/\rho$ terms.
This implies that the flat space limit
of the gravitational vertex for $(A)dS$ massless field
in $d=4$ is less singular that the
one in $d>4$.

\noindent
{\bf iii)} We note that,
for the massless spin 5/2 field and the massless spin 2 field in $(A)dS_d$,
there are no higher-derivative vertices subject to the condition $k_{max}\leq 3$.
This implies that the gravitational vertex for the
massless spin 5/2 field in $(A)dS_d$ is unique.

\noindent
{\bf iv)}
Reason for choice of the scheme
with $m_{3,1}(1) = 0$, $m_{3,3}(1)= 0$
can easily be understood by noting that the
{\it free} massive spin 5/2 field $\psik$ taken
in the massless limit
is decomposed into two decoupling systems
- one massless spin 5/2 field $\psi_2$
and one massive spin 3/2 field described by the fields
$\psi_1$, $\psi_0$.
The coefficients $m_{3,1}(1)$, $m_{3,3}(1)$ describe interactions
of the massive spin 3/2 field and the massless spin 2 field (see
\rf{LLthreedef01}). Therefore we set these
coefficients
equal to zero while solving the equations for the gravitational
vertex of the massless spin 5/2 field.

\subsection{ Gravitational vertex
for partial massless spin 5/2 field in $(A)dS_d$, $d>4$}

We note that the partial massless field is described by
fields $\psi_2$ and $\psi_1$ \rf{psisimnot01} with the mass
parameter given in \rf{caseads3}.
Solution to the equations for
$m_{a,b}$, $m_{a,b}^\im$
required by on-shell gauge invariance can uniquely be determined
by imposing the conditions $m_{3,2}=0$, $m_{3,3}(1)=0$, $m_{3,3}^\im(1)=0$.
In this way, we obtain
\beq
\label{parmasslmm01} && m_{2,1}(0) =
\frac{1}{4(d-3)\sqrt{-\coscon}}\,,
\\[3pt]
&& m_{2,1}(1) = -\frac{d^3 -4d^2 + d
+8}{4d(d-3)(d^2-1)\sqrt{-\coscon}}\,,
\\[3pt]
&& m_{2,2} = \frac{1}{(d-3)\sqrt{-\coscon}} \,,
\\[3pt]
&& m_{2,3} = -\frac{d^2-2d-4}{2d(d-3)\fwt_1(0)}  \,,
\\[3pt]
&& m_{3,1}(0) =0\,,
\\[3pt]
&& m_{3,1}(1) = -\frac{2}{d(d-3)\fwt_1^2(0)}\,,
\\[3pt]
&& m_{3,2} =  0 \,,
\\
&& m_{3,3}(0) = -\frac{1}{4(d-3)\fwt_1(0)\sqrt{-\coscon}}\,,
%
\\
&& m_{3,3}(1) = 0\,,
\\
&& m_{3,4} = -\frac{1}{(d-3)\fwt_1(0)\sqrt{-\coscon}}\,,
\\
\label{parmasslmm11} && m_{3,5} = 0 \,,
\\
\label{parmasslmm12}
&& m_{a,b}^\im = 0\,, \qquad \hbox{ for all } \  a,b\,.
\eeq
Adding the minimal gravitational vertex $\LL_\smone$ to
the sum $\LL_\smtwo+\LL_\smthree$ with above-given $m_{ab}$, $m_{a,b}^\im$
and ignoring contribution of the field $\psi_0$ to $\LL_\smone$
gives the gravitational vertex for the partial massless spin 5/2 field.
We note that the mass parameter $\mas$ for partial massless spin 5/2 field
given in \rf{caseads3} implies
\be \label{f0forparmass01}
\fwt_1(0)= \Bigl(\frac{d^2-1}{d} \coscon \Bigr)^{1/2}\,.\ee
The following remarks are in order.

\noindent
{\bf i)} Since $m_{2,a}\ne0$, $m_{3,a}\ne0$ for some $a$, we see that
the gravitational vertex
involves, in addition to the minimal gravitational vertex,
contributions with two and three derivatives,
i.e., the gravitational vertex has $k_{min}=1$, $k_{max}=3$. When $d>4$,
it is not possible to build gravitational vertex
with $k_{max}=2$.

\noindent
{\bf ii)} From
\rf{parmasslmm01}-\rf{f0forparmass01}, we see that, in view of
$1/\sqrt{\coscon}$ and $1/\coscon$ factors, the
gravitational vertex for the partial massless field in $(A)dS$
does not have smooth
flat space limit, $ \coscon \rightarrow 0$.
Below, we demonstrate that, when $d=4$, the
gravitational vertex does not involve three-derivative
contributions, i.e., $1/\rho$ terms contribute to the gravitational vertex
of the partial massless field only
when $d>4$.

\noindent
{\bf iii)}
As will be explained below, for the partial massless spin 5/2 field,
there is only one
higher-derivative vertex having
$k_{min}=2$ and $k_{max}=3$. Due to this, there is one parametric freedom in
choosing the gravitational vertex of the partial massless spin 5/2 field,
i.e., the restrictions
imposed by gauge invariance do not allow
to determine the gravitational vertex uniquely.
Therefore we need some additional restriction
which allows us to determine the gravitational vertex uniquely.
We chose the additional restriction to be $m_{3,2}=0$, which amounts to
requiring that $\psi_2 C D \psi_2$-terms (see \rf{LLthreedef01})
do not contribute to the gravitational vertex. We note that,
in analysis of the gravitational vertex taken to cubic order in fields,
the particular
choice of $m_{3,2}$ amounts just to choice of scheme for
the gravitational vertex. Physical values
of $m_{3,2}$ can be determined by study of complete,
i.e., to all order in fields, theory of massive higher-spin fields.

\noindent
{\bf iv)} Recall that the scheme we chose is specified not only
by the condition $m_{3,2}=0$
but also by the conditions $m_{3,3}(1) = 0$, $m_{3,3}^\im(1)= 0$.
Reason for the latter conditions
can easily be understood by noting that the {\it free} massive spin 5/2
field $\psik$ taken in
the partial massless limit
is decomposed into two decoupling systems -
one partial massless spin 5/2 field, which is
described by the fields $\psi_2$, $\psi_1$,
and one massive spin 1/2 field $\psi_0$.
The coefficients $m_{3,3}(1)$, $m_{3,3}^\im(1)$ describe interactions
of the partial massless spin 5/2 field with the massive spin 1/2 field
and the massless spin 2 field (see Section \ref{exover01}).
Therefore we set these coefficients
equal to zero while solving the equations for the gravitational
vertex of the partial massless spin 5/2 field.

\noindent
{\bf v)}
Because
some of the coefficients $m_{a,b}$ are complex-valued,
the gravitational vertex also turns out to be complex-valued. Appearance
of complex-valued Lagrangian for interacting partial massless field
was expected.
This is to say that even the mass part of the free Lagrangian for the partial
massless field is complex-valued.
We note that the key point is not positivity or even
reality of the mass part of action, but rather stability
of the energy and unitarity of the underlying physical
representations. For bosons a negative mass term is allowed in $AdS$
(the Breitenlohner-Freedman bound),
while partially massless fermions even have an imaginary mass term in
their actions but are still stable and unitary in $dS$. Partially
massless fermions are not unitary in $AdS$ (see
Refs.\cite{Deser:2001xr,Deser:2001us}).
Physical applications of partial massless fields are still to
be understood.

\subsection{ Higher-derivative vertex for partial massless spin 5/2 field and
massless spin 2 in $(A)dS_d$, $d > 4$}\label{subsechigderverparmas01}

For partial massless
spin 5/2 field and massless spin 2 field, there is only one
higher-derivative vertex subject to the condition
$2\leq k_{min}\leq k_{max}\leq 3$.
The vertex takes
$k_{min}=2$, $k_{max}=3$ and we now discuss this vertex.

Solution to the equations for $m_{a,b}$, $m_{a,b}^\im$
required by on-shell gauge invariance can uniquely be determined
by imposing the conditions $m_{3,3}(1)=0$, $m_{3,3}^\im(1)=0$
and by equating the gravitational constant $\kappa$
to zero.
In this way, we obtain
\beq
\label{parhighdera01int} && m_{2,1}(0) = -\frac{1}{4\sqrt{-
\coscon}}\, g\,,
\\[3pt]
&& m_{2,1}(1) =  \frac{3d^3-7d^2-2d+12}{4d^2(d+1)\sqrt{-\coscon}}g\,,
\\[3pt]
&& m_{2,2}= -\frac{1}{\sqrt{- \coscon }}\, g\,,
\\[3pt]
&& m_{2,3}= \frac{ (d^2 + d -1) \fwt_1(0) }{d (d+1) \coscon }\, g\,,
\\[3pt]
&& m_{3,1}(0) = \frac{1}{\coscon }\, g\,,
\\[3pt]
&& m_{3,1}(1) = \frac{d^2-d+2}{d(d+1) \coscon }\, g \,,
\\[3pt]
&& m_{3,2} =  \frac{1}{ \coscon }\, g \,,
\\[3pt]
&& m_{3,3}(0) = \frac{d-1}{2  d   \fwt_1(0) \sqrt{- \coscon } } \, g
\,,
\\[3pt]
&& m_{3,3}(1) = 0\,,
\\[3pt]
&& m_{3,4} = \frac{2(d-1)}{ d \fwt_1(0) \sqrt{- \coscon }  }\, g\,,
\\[3pt]
\label{parhighdera11int} && m_{3,5} = 0\,,
\\
\label{parhighdera12int}
&& m_{a,b}^\im = 0\,, \qquad \hbox{ for all } \  a,b\,,
\eeq
where $\fwt_1(0)$ is given in \rf{f0forparmass01},
and $g$ stands for a coupling constant
of the partial massless spin 5/2 field to
the massless spin 2 field.

The following remarks are in order.

\noindent
{\bf i)} Since $m_{2,a}\ne0$, $m_{3,a}\ne0$ for some $a$, we see that the
higher-derivative vertex is indeed
involves contributions with two and three derivatives, i.e.,
$k_{min}=2$, $k_{max}=3$.

\noindent
{\bf ii)}
We note that the vertex
under consideration is non-trivial only when $d>4$, i.e.,
the vertex does not survive reduction to $d=4$. This, using relations
\rf{sec100-01}-\rf{sec100-01n001}
from Appendix F,
one can make sure that two-derivative terms and
three-derivative terms
\rf{parhighdera01int}-\rf{parhighdera12int} cancel when $d=4$.

\noindent
{\bf iii)}
Reason for choice of the scheme
with $m_{3,3}(1) = 0$, $m_{3,3}^\im(1)= 0$
is that the {\it free} massive spin 5/2 field $\psik$
taken in the partial massless limit
is decomposed into two decoupling systems - one partial massless spin 5/2 field,
which is described by the fields $\psi_2$, $\psi_1$,
and one massive spin 1/2 field $\psi_0$.
The coefficients $m_{3,3}(1)$, $m_{3,3}^\im(1)$ describe higher-derivative
interactions
of the partial massless spin 5/2 field with the massive
spin 1/2 field and the massless spin 2 field (see Section \ref{exover01}).
Therefore we set these coefficients
equal to zero while solving the equations for higher-derivative
vertex of the partial massless spin 5/2 field.

\noindent
{\bf iv)}
Some of the coefficients $m_{a,b}$ are complex-valued and hence
the higher-derivative vertex is also complex-valued.
Since even the mass part of the free Lagrangian for the partial
massless field is complex-valued, appearance
of a complex-valued higher-derivative vertex for the partial massless field
was expected. Physical applications of the complex-valued vertex are still to
be understood.

\noindent
{\bf v)}
As we have said in Section \ref{hihdersubsubsec}, though the two
higher-derivative vertices for the massive spin 5/2 field
are singular in the partial massless limit,
$\fwt_1(1)\rightarrow 0 $,
one can build some special combination of those two higher-derivative vertices
which has smooth partial massless limit. We now demonstrate how this
comes about.
Using the notation $m_{a,b}^{ {\bf A} }$ for
$m_{a,b}$ given in \rf{highdera01}-\rf{highdera11}, and
$m_{a,b}^{{\bf B}}$ for $m_{a,b}$ given in
\rf{highderb01}-\rf{highderb11} we introduce new $m_{a,b}$ by
\be \label{mabparmassmolom01}
m_{a,b} = m_{a,b}^{\bf A} + \frac{2(d-2)g}{g'} m_{a,b}^{\bf B}\,, \ee
while $m_{a,b}^\im$ remain unchanged.
Making use of \rf{highdera01}-\rf{highdera11} and
\rf{highderb01}-\rf{highderb11} one can make sure that, firstly,
all $1/(\fwt_1(1)\rightarrow 0)$
singularities cancel in \rf{mabparmassmolom01} and, secondly, the coefficients
$m_{a,b}$ given \rf{mabparmassmolom01} coincide with $m_{a,b}$
given in \rf{parhighdera01int}-\rf{parhighdera11int} as $\fwt_1(1)\rightarrow 0$.

\subsection{ Higher-derivative vertex for massless spin 5/2
field in flat space, $d>4$}\label{mas5mas5mas2dgr4}

In double limit, i.e., the flat space limit, $\rho \rightarrow 0$,
and the massless field limit, $\mas \rightarrow 0$,
the {\it free} massive spin 5/2 field $|\psi\rangle$ \rf{psikexp01} is
decomposed into three decoupling systems -- one massless spin 5/2 field
$\psi_2$, one massless spin 3/2 field $\psi_1$,  and one massless spin
1/2 field $\psi_0$. We find one
higher-derivative cubic vertex (with $k_{max} \leq 3$) that describes
interaction like $\psi_2-\psi_2-$(massless spin 2 field).
Equating the gravitational constant $\kappa$ to zero we obtain
\beq \label{552ver01flatd}
&& m_{3,1}(0) = g \,,\qquad m_{3,2} = g\,,
\\
\label{552ver02flatd}
&& \hbox{ all remaining $m_{a,b}$ and $m_{a,b}^\im$ are equal to zero},
\eeq
where $g$ stands for a coupling constant. From these expressions, we see
that the vertex has $k_{min}=k_{max}=3$.
We also note that this vertex is non-trivial only when $d>4$.

The following remarks are in order.

\noindent
{\bf i}) As we have learned earlier,
in $(A)dS_d$, the massless spin 5/2 field does not have higher-derivative
vertices subject to the condition $k_{max}\leq 3$. This implies that the
{\it higher-derivative vertex} of the massless spin 5/2 field in flat space,
\rf{552ver01flatd},\rf{552ver02flatd}, does not
have its counterpart in $(A)dS_d$ space. However, it turns out
that the {\it higher-derivative vertex of the massless spin 5/2
field in flat space}
can be related to the {\it gravitational
vertex of the massless spin 5/2 field} in $(A)dS_d$%
\footnote{We thank M.A.Vasiliev for pointing out this.}.
This, let $(\LL_\smone
+ \LL_\smtwo + \LL_\smthree )^{grav, (A)dS_d}$
be the gravitational vertex of the massless spin 5/2 field in $(A)dS_d$, where
$\LL_\smone$ is the minimal gravitational vertex, while
$\LL_\smtwo$ and $\LL_\smthree$ are the respective
two-derivative and three-derivative  vertices given in
\rf{masslmm01}-\rf{masslmm12}. Let $\LL_\smthree^{high-der. flat}$
be the higher-derivative vertex given in \rf{552ver01flatd},\rf{552ver02flatd}.
Then,
taking the gravitational vertex with $\kappa=1/\sqrt{2}$, we obtain the relation
\be
\lim_{\rho \rightarrow 0}  \rho  (\LL_\smone
+ \LL_\smtwo + \LL_\smthree )^{grav, (A)dS_d}
=
\frac{1}{2(d-1)g} \LL_\smthree^{high-der. flat}\,.
\ee
This relation implies that certain higher-derivative vertices in flat space
can be obtained from gravitational vertices in $(A)dS_d$.

\noindent
{\bf ii}) We note that
the {\it higher-derivative vertex for massless field in flat space}
\rf{552ver01flatd},\rf{552ver02flatd} can be related to the
{\it higher-derivative
vertex for the partial massless field} in $(A)dS_d$
given in Section \ref{subsechigderverparmas01}.
This, let $(\LL_\smtwo + \LL_\smthree )_{part.mass}^{high-der, (A)dS_d}$
be the higher-derivative vertex for the partial massless spin 5/2 field
in $(A)dS_d$, where
$\LL_\smtwo$ and $\LL_\smthree$ are the respective
two-derivative and three-derivative  vertices given in
\rf{parhighdera01int}-\rf{parhighdera12int}. Let $\LL_\smthree^{high-der. flat}$
be the higher-derivative vertex given in \rf{552ver01flatd},\rf{552ver02flatd}.
Then, ignoring contributions of the fields $\psi_1$ and $\psi_0$ to the
higher-derivative vertex of the partial massless field, we obtain the relation
\be
\lim_{\rho \rightarrow 0}
\rho  ( \LL_\smtwo
+ \LL_\smthree )_{part.mass;\,\psi_1=\psi_0=0}^{high-der, (A)dS_d}
=
\LL_\smthree^{high-der. flat}\,.
\ee

\newsection{ Uniform gravitational vertex for massive, massless
and partial massless spin 5/2 fields in $(A)dS_d$, $d>4$}\label{secuniver01}

Gravitational
vertex for the massive spin 5/2 field given in \rf{gravmassiv01}-\rf{gravmassiv12}
does not have smooth massless and partial massless limits. However by
adding to this vertex the suitable combination of the higher-derivative
vertices given
in \rf{highdera01}-\rf{highdera11n1}, \rf{highderb01}-\rf{highderb11n1}
we can obtain new gravitational vertex that has
smooth massless and partial massless limits. We now demonstrate this.

Using the notation $m_{a,b}^{grav}$ for $m_{a,b}$ given in
\rf{gravmassiv01}-\rf{gravmassiv11}, $m_{a,b}^{ {\bf A} }$ for
$m_{a,b}$ given in \rf{highdera01}-\rf{highdera11}, and
$m_{a,b}^{{\bf B}}$ for $m_{a,b}$ given in
\rf{highderb01}-\rf{highderb11}, we introduce new $m_{a,b}$ by

\be \label{newmab01}
m_{a,b} = m_{a,b}^{grav} + \omega \Bigl( \frac{1}{g}m_{a,b}^{
{\bf A} } - \frac{2d}{g'}m_{a,b}^{{\bf B}}\Bigr),
\ee
where
\be \omega \equiv \frac{\mas^2}{12\mas^2 - d(d-4) \coscon }\,,
\ee
and $m_{a,b}^\im$ remain unchanged.
New $m_{a,b}$ \rf{newmab01} take the form
\beq
\label{gravversmoo01} && m_{2,1}(0) = \frac{\omega}{\mas}\,,
\\[3pt]
&& m_{2,1}(1) = \frac{(d+2)\omega}{d\mas}\,,
\\[3pt]
&& m_{2,2} =  \frac{4\omega}{\mas}\,,
\\[3pt]
&& m_{2,3} = -\frac{6 \fwt_1(0) \omega }{\mas^2} \,,
\\[3pt]
&& m_{3,1}(0) = -\frac{2d}{\mas^2}\omega\,,
\\[3pt]
&& m_{3,1}(1) = -\frac{2(d-2) \omega }{\mas^2}\,,
\\[3pt]
&& m_{3,2} = -\frac{2d \omega }{\mas^2}\,,
\\
&& m_{3,3}(0) =0\,, \qquad
m_{3,3}(1) =0\,,
\\
\label{gravversmoo11} &&  m_{3,4} = 0\,, \qquad  m_{3,5} =0\,,
\\
&& m_{a,b}^\im =0\,, \quad \hbox{ for all } \ a,b\,.
\eeq
Taking into account values of the mass parameter $\mas$
for the massless and partial massless
fields, \rf{caseads2},\rf{caseads3}, we see that the new gravitational
vertex has smooth
massless and partial massless limits. We note that
choice of gravitational vertex having smooth massless and partial
limit is not unique. One can expect that gravitational vertices for
massive fields that have smooth massless limit may be of interest in
the context of the tensionless limit of superstring theory.

\newsection{ Massive, massless and partial massless
spin 5/2 fields, $d=4$}\label{secmasverd401}

\subsection{ Gravitational vertex for massive spin 5/2 field in $(A)dS_4$}

We now discuss the gravitational vertex for
the massive Dirac spin 5/2 field in $d=4$. We note that all vertices obtained
above for $d>4$ are valid for $d=4$ too. But, when $d=4$,
vertex operators \rf{M21def}-\rf{M35cdef}
do not constitute basis of independent
vertex operators. This, when $d=4$, the three-derivative
vertex operators $M_{3,1}$, $M_{3,2}$, $M_{3,4}$, $M_{3,4}^\im$
can be expressed in terms of the remaining three-derivative and
two-derivative vertex operators (for details, see Appendix F ).
We still use vertex operators \rf{M21def}-\rf{M35cdef}
for presentation of our results.
In $d=4$, the higher-derivative contributions, \rf{gravmassiv01}-\rf{gravmassiv12},
to the gravitational vertex take the form%
\footnote{ In formulas for the gravitational vertex given
in Section \ref{subsubgravvermass01},
we plug the representation for $M_{3,1}$, $M_{3,2}$, $M_{3,4}$
in terms of the remaining three-derivative and
two-derivative vertex operators and express the gravitational vertex entirely
in terms of those vertices that form basis of $4d$ vertices. The
fact that the vertex operators $M_{3,1}$, $M_{3,2}$, $M_{3,4}$ do not enter into
basis of $4d$ vertex operators
manifests itself in zero values of $m_{3,1}$, $m_{3,2}$, $m_{3,4}$
for $4d$ gravitational vertex. For discussion of base vertex operators in $d=4$,
see Appendix F.}
\beq
\label{grav4d01} && m_{2,1}(0) = 0\,,
\\[3pt]
&& m_{2,1}(1) = \frac{3\mas}{4\fwt_1^2(0)}\,,
\\[3pt]
&& m_{2,2} = 0\,,
\\[3pt]
&& m_{2,3} = -\frac{1}{\fwt_1(0)} \,,
\\[3pt]
&& m_{3,1}(0) = 0\,,
\\[3pt]
&& m_{3,1}(1) =0\,,
\\[3pt]
&& m_{3,2} =  0 \,,
\\[3pt]
&& m_{3,3}(0) =0\,,
\\[3pt]
&& m_{3,3}(1) = - \frac{3\mas}{8\fwt_1^2(0)\fwt_1(1)}\,,
\\[3pt]
&& m_{3,4} = 0\,,
\\[3pt]
\label{grav4d11} && m_{3,5} = \frac{1}{2\fwt_1(0)\fwt_1(1)} \,,
\\[3pt]
\label{grav4d12} && m_{a,b}^\im = 0\,,\qquad \hbox{ for all } \ a,b\,.
\eeq
Since some $m_{3,a}\ne 0$, the gravitational vertex
still involves three-derivative
contributions. But, in contrast
to the gravitational vertex for $d>4$, these contributions
can be canceled by adding higher-derivative
vertex to the gravitational vertex.
Before doing so, let us consider higher-derivative
vertex for the massive spin 5/2 field in $d=4$.

\subsection{Higher-derivative vertex for massive spin 5/2 field and massless
spin 2 field in $(A)dS_4$}

Higher-derivative vertices for the massive spin 5/2
field in $(A)dS_4$ can be obtained  by allowing $d=4$ in the
two higher-derivative vertices (see \rf{highdera01}-\rf{highdera11n1}
and \rf{highderb01}-\rf{highderb11n1})
obtained for $d>4$.
It turns out that the latter two
vertices become identical in $d=4$. In other words,
when $2\leq k_{min}\leq k_{max}\leq 3$ and $d=4$, there
is only one higher-derivative vertex. We obtain this vertex by
allowing $d=4$ in
\rf{highdera01}-\rf{highdera11n1} (vertex given in
\rf{highderb01}-\rf{highderb11n1} takes the same form by module of
overall normalization factor).

When $d=4$, vertex given in
\rf{highdera01}-\rf{highdera11n1} takes the form
\beq
\label{higder4d01} && m_{2,1}(0) = \frac{1}{\mas} g \,,
\\[3pt]
&& m_{2,1}(1) = - \frac{3\mas}{\fwt_1^2(0)} g \,,
\\[3pt]
&& m_{2,2} =  \frac{4}{\mas}g\,,
\\[3pt]
&& m_{2,3} = \frac{4}{\fwt_1(0)} g  \,,
\\[3pt]
&& m_{3,1}(0) = 0\,,
\\[3pt]
&& m_{3,1}(1) = 0  \,,
\\[3pt]
&& m_{3,2} =  0 \,,
\\[3pt]
&& m_{3,3}(0) = 0 \,,
\\[3pt]
&& m_{3,3}(1) = \frac{3\mas}{2\fwt_1^2(0)\fwt_1(1) } g \,,
\\[3pt]
&& m_{3,4} = 0\,,
\\[3pt]
\label{higder4d11} && m_{3,5} = -\frac{2}{\fwt_1(0)\fwt_1(1)} g \,,
\\
\label{higder4d12}  && m_{a,b}^\im=0\,,\qquad \hbox{ for all } \ a,b.
\eeq
Since $m_{2,a}\ne0$, $m_{3,a}\ne0$ for some $a$, the vertex involves
contributions with two and three derivatives.

The higher-derivative vertex under consideration
is singular in the limit of massless field,
$\fwt_1(0)\rightarrow 0 $,
and the limit of partial massless field, $\fwt_1(1)\rightarrow 0 $.
This implies that there are no higher-derivative vertices%
\footnote{ In \rf{higder4d01}-\rf{higder4d12},
we can formally re-scale the coupling constant,
$g \rightarrow \fwt_1^2(0) g$, to get well defined vertex in the limit
of massless field. In doing so, we obtain non-trivial
values only for $m_{2,1}(1)$, $m_{3,3}(1)$ in the massless limit
$\fwt_1(0)\rightarrow 0 $. But such vertex describes interaction
of spin 3/2 field with the massless spin 2 field.
Re-scaling $g \rightarrow \fwt_1(1) g $ and taking limit
$\fwt_1(1)\rightarrow 0 $ leads to vertex describing
interaction of the partial massless spin 5/2 field with
massive spin 1/2 field and the massless spin 2 field in $(A)dS_4$.
This vertex is discussed in Section \ref{exover01}.},
for both the massless spin 5/2 field and the partial massless spin 5/2
field in $(A)dS_4$ when $k_{max}\leq 3$.

\subsection{ Uniform
gravitational vertex for massive, massless and partial
massless field in $(A)dS_4$}\label{secuniver02}

Comparing two-derivative and three-derivative terms
\rf{grav4d01}-\rf{grav4d12} of the
gravitational vertex
with the higher-derivative vertex in \rf{higder4d01}-\rf{higder4d12},
we see that all three-derivative
terms of the gravitational vertex can be canceled by adding
the higher-derivative vertex
(with appropriate choice of the overall normalization factor)
to the gravitational vertex. We now
demonstrate this. Using the notation $m_{a,b}^{grav}$ for
$m_{a,b}$ given in \rf{grav4d01}-\rf{grav4d11}, and $m_{a,b}^{high-der}$
for $m_{a,b}$ given in \rf{higder4d01}-\rf{higder4d11}, we
introduce new $m_{a,b}$ by
\be
m_{a,b} = m_{a,b}^{grav} + \frac{1}{4g} m_{a,b}^{high-der }\,,
\ee
while $m_{a,b}^\im$ remain unchanged. The new $m_{a,b}$ take the form
\beq
\label{grav4d01sh} && m_{2,1}(0) = \frac{1}{4\mas}\,, \qquad
m_{2,2} = \frac{1}{\mas}\,,
\\[3pt]
\label{grav4d0501sh}
&& \hbox{ all remaining $m_{a,b}$ and $m_{a,b}^\im$ are equal to zero}.
\eeq
We see that all three-derivative terms governed by $m_{3,a}$
cancel, i.e., $\LL_\smthree=0$. This implies that the gravitational
vertex involves, in addition to the
minimal vertex, only two-derivative contributions governed by $\LL_\smtwo$
with above-given $m_{2,a}$.

A few remarks are in order.

\noindent
{\bf i)} The vertex \rf{grav4d01sh},\rf{grav4d0501sh} has smooth limits to
the massless and partial massless fields in $(A)dS_4$. Taking these
limits, we obtain the respective
gravitational vertices for the massless and partial massless fields in $(A)dS_4$.
This, to obtain the gravitational vertices for the massless and partial massless
fields we should simply plug the respective
limiting values of mass parameter, $\mas = 2\sqrt{- \coscon }$ and
$\mas = \sqrt{- \coscon }$, in \rf{grav4d01sh},\rf{grav4d0501sh}%
\footnote{ Obviously, for the case of massless field, we
should drop contributions of the fields $\psi_1$, $\psi_0$
to the minimal gravitational vertex $\LL_\smone$,
while for the case of partial massless field, we
should drop contributions of the field $\psi_0$
to $\LL_\smone$.}.
In other words, the vertex \rf{grav4d01sh},\rf{grav4d0501sh}
provides uniform description of gravitational interaction
of the massive, massless, and partial massless spin 5/2 fields in $(A)dS_4$.

\noindent
{\bf ii)} The vertex \rf{grav4d01sh},\rf{grav4d0501sh} taken
for massive field, $m\ne 0$, has smooth flat
space limit, $\rho\rightarrow 0$.
In this limit, we obtain the gravitational vertex
for the massive spin 5/2 field in $4d$ flat space \cite{Porrati:1993in}.
In other words, in contrast to the massless and partial massless fields,
the gravitational interaction of
the massive spin 5/2 field in $(A)dS$ space has smooth limit
to the gravitational interaction of the massive spin 5/2 field in {\it flat} space.

\noindent
{\bf iii)}
To restore dependence on the
gravitational constant $\kappa$, we should simply multiply the r.h.s.
of relations \rf{grav4d01sh},\rf{grav4d0501sh} by $\sqrt{2}\kappa$.

\subsection{Higher-derivative vertex for massless spin 5/2 field and
massless spin 2 in flat space, $d=4$}

As we have said, in double limit, i.e., the flat space limit, $\rho \rightarrow 0$,
and the massless field limit, $\mas \rightarrow 0$,
the {\it free }massive spin 5/2 field $|\psi\rangle$ \rf{psikexp01} is
decomposed into three decoupling systems -- one massless spin 5/2 field $\psi_2$,
one massless spin 3/2 field $\psi_1$,  and one massless spin
1/2 field $\psi_0$. We find one
higher-derivative cubic vertex (with $k_{max} \leq 3$) that describes
interaction like
$\psi_2-\psi_2-$(massless spin 2 field).
Equating the gravitational constant $\kappa$ to zero we obtain
\beq \label{hihderkmin2kmax201}
&& m_{2,1}(0) = g \,,\qquad m_{2,2} = 4g\,,
\\
\label{hihderkmin2kmax202}
&& \hbox{ all remaining $m_{a,b}$ and $m_{a,b}^\im$ are equal to zero},
\eeq
where $g$ stands for a coupling constant.
We see that the vertex has $k_{min}=k_{max}=2$.
Recall, in $(A)dS_4$, the massless spin 5/2 field does not have higher-derivative
vertices subject to the condition $k_{max}\leq 3$. This implies that the
{\it higher-derivative vertex} of the massless spin 5/2 field in $4d$ flat space,
\rf{hihderkmin2kmax201},\rf{hihderkmin2kmax202},
does not have its counterpart in $(A)dS_4$ space.
However,  by analogy with discussion in Section \ref{mas5mas5mas2dgr4},
the {\it higher-derivative vertex of the massless spin 5/2 field in $4d$
flat space},
\rf{hihderkmin2kmax201},\rf{hihderkmin2kmax202},
can be related to the {\it gravitational
vertex of the massless spin 5/2 field} in $(A)dS_4$.
This, let $(\LL_\smone
+ \LL_\smtwo)^{grav, (A)dS_4}$
be the gravitational vertex of the massless spin 5/2 field in $(A)dS_4$, where
$\LL_\smone$ is the minimal gravitational vertex, while
$\LL_\smtwo$ is the two-derivative vertex given in
\rf{grav4d01sh},\rf{grav4d0501sh}.
Let $\LL_\smtwo^{high-der. flat}$
be the higher-derivative vertex given in
\rf{hihderkmin2kmax201},\rf{hihderkmin2kmax202}.
Then, taking the gravitational vertex
with $\kappa=1/\sqrt{2}$, we obtain
\be
\lim_{\rho \rightarrow 0}  \sqrt{-\coscon}   (\LL_\smone
+ \LL_\smtwo  )^{grav, (A)dS_4}
=
\frac{1}{8g} \LL_\smtwo^{high-der. flat}\,,
\ee
where we use relation $\mas=2\sqrt{-\coscon}$, \rf{caseads2},
for mass parameter of massless field when $d=4$.

\newsection{Higher-derivative vertices for (partial) massless spin 5/2 field,
(massive) massless spin 1/2 field and
massless spin 2 field}\label{speintver01}

The massless and partial massless limits lead to some special
higher-derivative vertices for spin 5/2 field.
These special vertices describe interactions of spin 5/2 field
with low spin 3/2 and 1/2 fields and massless spin 2 field.
The vertices are listed in Table II. We now discuss them in turn.

\subsection{Higher-derivative vertices for partial massless spin 5/2 field,
massive spin 1/2 field and
massless spin 2 in $(A)dS_d$, $d \geq 4$}\label{exover01}

In the partial massless limit $\fwt_1(1)\rightarrow 0$, \rf{caseads3},
the {\it free} massive spin 5/2 field $|\psi\rangle$ \rf{psikexp01} is
decomposed into two decoupling systems -- one partial massless spin 5/2 field,
which is described by the fields $\psi_2$, $\psi_1$, and one massive spin
1/2 field $\psi_0$. The mass parameter of the partial
massless field is given in \rf{caseads3}, while the mass parameter
of the field $\psi_0$ is equal to $\mas = \frac{d+2}{2}\sqrt{-\coscon}$,
\cite{Metsaev:2006zy}.
We find two
higher-derivative cubic vertices (with $k_{max} \leq 3$) that describe
interactions like
$(\psi_2, \psi_1)-\psi_0-$(massless spin 2 field).
Both these vertices are nontrivial for $d\geq 4$ and
take $k_{min}=k_{max}=3$. We now discuss these vertices
in turn.

\medskip
\noindent {\bf a) First higher-derivative vertex}.
For this case, solution to the equations for $m_{a,b}$, $m_{a,b}^\im$
required by on-shell
gauge invariance can uniquely be determined by imposing
the conditions $m_{3,2}=0$,
$m_{3,4}=0$,
and by equating the gravitational constant and $m_{a,b}^\im$ to zero.
In this way, we obtain
\beq
\label{aparhighdera01int}
&& m_{2,a} = 0\,, \qquad \hbox{ for all} \ a\,,
\\
&& m_{3,1}(0) = 0\,,
\qquad
m_{3,1}(1) = 0\,,
\\
&& m_{3,2} = 0\,,
\\
&& m_{3,3}(0) = 0\,,
\\
&& m_{3,3}(1) = \frac{d-1}{ d \coscon }\, g' \,,
\\
&& m_{3,4} = 0\,,
\\
\label{aparhighdera11int} && m_{3,5} = -\frac{\fwt_1(0)}{ \coscon
\sqrt{- \coscon }} \, g' \,,
\\[3pt]
\label{aparhighdera11int01} && m_{a,b}^\im = 0\,, \hbox{ for all } \
a,b\,,
\eeq
where $g'$ stands for a coupling constant and $\fwt_1(0)$ is given in
\rf{f0forparmass01}.

\medskip
\noindent {\bf b) Second higher-derivative vertex}.
This vertex is governed entirely by the vertex operators $M_{a,b}^\im$.
This to say that, firstly, we equate
the gravitational constant and $m_{a,b}$ to zero and, secondly, we find
that
solution to the equations for $m_{a,b}^\im$ imposed by on-shell
gauge invariance is uniquely determined.
We obtain solution%
\footnote{ We note that only for the partial massless field (and massless field in
flat space) requirement of gauge invariance allows nontrivial
$m_{a,b}^\im$.}%
\beq
&& m_{a,b}=0\,,\qquad \hbox{ for all } \ a,b\,,
\\
\label{parhighdera01intcc} && m_{2,3}^\im =  0\,,
\\
&& m_{3,3}^\im(0) = 0\,,
\\
&& m_{3,3}^\im(1) = \frac{d-1}{d \coscon } g^\im{}'\,,
\\
&& m_{3,4}^\im = 0\,,
\\
\label{parhighdera11intcc} && m_{3,5}^\im = -
\frac{\fwt_1(0)}{\coscon\sqrt{-\coscon}} g^\im{}'\,,
\eeq
where $g^\im{}'$ stands for a coupling constant and $\fwt_1(0)$ is given in
\rf{f0forparmass01}.

\subsection{Higher-derivative vertices for massless spin 5/2 field,
massless spin 3/2,\, 1/2 fields, and
massless spin 2 in flat space, $d \geq 4$}\label{exover01nn1}

In double limit, i.e., the flat space limit, $\rho \rightarrow 0$,
and the massless field limit, $\mas \rightarrow 0$,
the {\it free} massive spin 5/2 field $|\psi\rangle$ \rf{psikexp01} is
decomposed into three decoupling systems -- one massless spin 5/2 field $\psi_2$,
one massless spin 3/2 field $\psi_1$,  and one massless spin
1/2 field $\psi_0$. We find two
higher-derivative cubic vertices (with $k_{max} \leq 3$) that describe
interactions like
$\psi_2-\psi_1-$(massless spin 2 field)
and two
higher-derivative cubic vertices (with $k_{max} \leq 3$) that describe
interactions like
$\psi_2-\psi_0-$(massless spin 2 field).
All these vertices have $k_{min}=k_{max}=3$. Equating the gravitational constant
$\kappa$ to zero, we obtain the following vertices.

\noindent
{\bf A}) Vertex $\psi_2-\psi_1-$(massless spin 2 field):
\beq \label{speintverA01}
&& m_{3,3}(0) = g \,,\qquad m_{3,4} = 4g\,,
\\
&& \hbox{ all remaining $m_{a,b}$ and $m_{a,b}^\im$ are equal to zero}.
\eeq

\noindent
{\bf B}) Vertex $\psi_2-\psi_1-$(massless spin 2 field):
\beq \label{speintverB01}
&& m_{3,3}^\im(0) = g^\im \,, \qquad m_{3,4}^\im = 4g^\im\,,
\\
&& \hbox{ all remaining $m_{a,b}$ and $m_{a,b}^\im$ are equal to zero}.
\eeq

\medskip

\noindent
{\bf C}) Vertex $\psi_2-\psi_0-$(massless spin 2 field):
\beq \label{speintverC01}
&& m_{3,5} = g' \,,
\\
&& \hbox{ all remaining $m_{a,b}$ and $m_{a,b}^\im$ are equal to zero}.
\eeq

\noindent
{\bf D}) Vertex $\psi_2-\psi_0-$(massless spin 2 field):
\beq \label{speintverD01}
&& m_{3,5}^\im = g^\im{}' \,,
\\
&& \hbox{ all remaining $m_{a,b}$ and $m_{a,b}^\im$ are equal to zero}.
\eeq
In \rf{speintverA01}-\rf{speintverD01},
$g$, $g^\im$, $g'$, and $g^\im{}'$ stand for coupling constants.
We note that the vertices ${\bf (A})$ and ${\bf (B})$ do not survive
reduction to $d=4$. In other words, the
vertices ${\bf (A})$ and ${\bf (B})$ are non-trivial only for
$d>4$.
We also note that the vertices ${\bf (C})$ and ${\bf (D})$
are non-trivial for both $d=4$ and $d>4$.

\newsection{On-shell gauge invariant formulation
of free massive spin 5/2 Dirac field and massless spin 2 field in
$(A)dS_d$}\label{ongauformsec01}

Because we are building interaction vertices in terms of the on-shell free
massive spin 5/2 gauge field and the massless spin 2 field
in $(A)dS$ we begin with discussion of formulation we use
to describe these fields%
\footnote{ There are various interesting approaches in the literature
which could be used to discuss interaction vertices. This is to say
that various Lagrangian formulations in terms of unconstrained fields
in flat space and $(A)dS$ space may be found e.g. in
\cite{Buchbinder:2004gp}-\cite{Francia:2002pt}. Application of BRST
approach to studying interacting fields in $AdS$ may be found in
\cite{Buchbinder:2006eq}.}.
On-shell description of the massless spin 2 field in $(A)dS$
is well known and is given at the end of
this Section. We begin with discussion of the massive spin 5/2 field.

To discuss Lorentz covariant and gauge invariant formulation of
the massive spin 5/2 field in $(A)dS_d$, we introduce
Dirac complex-valued tensor-spinor fields of
the $so(d-1,1)$ Lorentz algebra,
\be \label{collect} \psi^{AB\,\alpha},\quad \psi^{A \alpha}\,,\quad
\psi^{\alpha}\,, \ee
where $\psi^{AB\,\alpha}$ is symmetric in indices $A,B$. The
tensor-spinor fields $\psi^{AB\alpha}$, $\psi^{A\alpha}$
satisfy the $\gamma$-tracelessness  constraints
\be
\label{gamtracorig} \gamma^A \psi^{AB}=0\,,\qquad \gamma^A \psi^A
=0\,. \ee
We note that the fields $\psi^{AB\alpha}$, $\psi^{A\alpha}$, and $\psi^\alpha$
transform in the
respective spin 5/2, 3/2 and 1/2 non-chiral  representations
of the Lorentz algebra $so(d-1,1)$.

In order to obtain the gauge invariant description of the massive field
in an easy--to--use form we introduce a set of creation and
annihilation operators $\alpha^A$, $\zeta$ and $\bar{\alpha}^A$,
$\bar{\zeta}$ defined by the relations%
\footnote{ We use oscillator formulation
\cite{Lopatin:1987hz,Vasiliev:1987tk,Labastida:1987kw} to handle the
many indices appearing for tensor-spinor fields. It can also be
reformulated as an algebra acting on the symmetric-spinor bundle on
the manifold $M$ \cite{Hallowell:2005np}.
Note that the scalar oscillators $\zeta$, $\bar\zeta$ arise naturally
by a dimensional reduction \cite{Biswas:2002nk,Hallowell:2005np} from
flat space.
Our oscillators $\alpha^A$, $\bar\alpha^A$, $\zeta$, $\bar\zeta$ are
respective analogs of $dx^\mu$, $\partial_\mu$, $du$, $\partial_u$ of
Ref.\cite{Hallowell:2005np} dealing, among other thing, with massless
fermionic fields in $(A)dS$.}

\be\label{intver15}
[\bar\alpha^A,\,\alpha^B]= \eta^{AB}\,,\qquad
[\bar{\zeta},\,\zeta] = 1\,,
\qquad \bar\alpha^A|0\rangle=0\,,\qquad \bar{\zeta}|0\rangle=0\,,\ee
where $\eta^{AB}$ is the mostly positive flat metric tensor. The
oscillators $\alpha^A$, $\bar\alpha^A$ and $\zeta$, $\bar\zeta$
transform in the respective vector and scalar representations of the
$so(d-1,1)$ Lorentz algebra. The tensor-spinor fields \rf{collect}
can be collected into the ket-vector $|\psi\rangle$ defined by
\beq
\label{intver16n1}
&& |\psi\rangle  \equiv |\psi_2\rangle + \zeta |\psi_1\rangle +
\zeta^2 |\psi_0\rangle\,,
\\[3pt]
&& \label{genfun2} |\psi_2\rangle \equiv \psi^{AB\alpha}
\alpha^A\alpha^B |0\rangle\,,
\\[3pt]
 \label{genfun2nn1}
 && |\psi_1\rangle \equiv \psi^{A\alpha} \alpha^A |0\rangle\,,
\\[3pt]
\label{genfun2nn2}&& |\psi_0\rangle \equiv \psi^\alpha|0\rangle\,.
\eeq
Here and below, spinor indices of the ket-vectors are implicit.
The ket-vector
$|\psi\rangle$ \rf{intver16n1} satisfies the
algebraic constraints
\beq
&& \label{homcon2n} (N_\alpha + N_\zeta  - 2 )|\psi\rangle =0\,,
\\
&& \label{gamtra1n} \gamma\bar\alpha |\psi\rangle =0\,,
\\
\label{Nzetadef} && N_\alpha \equiv \alpha^A\bar\alpha^A\,,\qquad
N_\zeta \equiv \zeta\bar\zeta\,,
\qquad
\gamma\bar\alpha \equiv \gamma^A \bar\alpha^A\,.
\eeq
The constraint \rf{homcon2n} just tells us that the ket-vector
$|\psi\rangle$ is a degree-2
homogeneous polynomial in the oscillators $\alpha^A$ and $\zeta$,
while the constraint
\rf{gamtra1n} respects the $\gamma$-tracelessness constraints  for the
tensor-spinor fields $\psi^{AB}$
and $\psi^A$ (see \rf{gamtracorig}).

Gauge invariant equations of motion (E.o.M) for the massive spin 5/2 field
in $(A)dS_d$ space we found take the form
\be
\label{sec5006} \Bigl( \Dline + m_1 + 2\gamma\alpha \fwt_3
\bar\zeta\Bigr)\psik =0\,,
\ee
where $\gamma\alpha\equiv \gamma^A\alpha^A$. The
first-derivative Dirac operator $\Dline$ appearing in
E.o.M \rf{sec5006} is given by
\be \label{vardef01}
\Dline \equiv \gamma^A D^A\,,\qquad \qquad D^A = \eta^{AB}D_B\,,
\qquad \qquad
D_A \equiv e_A^\mu D_\mu\,,  \ee
where $e_A^\mu$ stands for inverse vielbein of $(A)dS_d$ space, while
$D_\mu$ stands for the Lorentz covariant derivative
\be \label{lorspiope} D_\mu \equiv
\partial_\mu
+\frac{1}{2}\omega_\mu^{AB}M^{AB}\,.\ee
The $\omega_\mu^{AB}$ is the Lorentz connection of $(A)dS_d$ space,
while a spin operator $M^{AB}$ forms a representation of the Lorentz
algebra $so(d-1,1)$:
\be\label{loralgspiope} M^{AB} = M_b^{AB} +
\frac{1}{2}\gamma^{AB}\,,\qquad M_b^{AB} \equiv \alpha^A \bar\alpha^B
- \alpha^B \bar\alpha^A\,,  \qquad \gamma^{AB} \equiv
\frac{1}{2}(\gamma^A\gamma^B - \gamma^B\gamma^A)\,.\ee
Operator $\fwt_3= \fwt_3(N_\zeta)$ and mass operator $m_1=
m_1(N_\zeta)$ appearing in  E.o.M \rf{sec5006} are given by
\beq
\label{mwt1sol} && m_1(N_\zeta)  = \frac{d + 2}{d + 2 - 2N_\zeta} \mas \,,
\\[5pt]
\label{Del3def}  && \fwt_3(N_\zeta)  = -\Bigl(\frac{d + 1 - N_\zeta}{ ( d -
2N_\zeta)^3}\, F(\mas, N_\zeta)\Bigr)^{1/2}\,,
\eeq
where the function $F(\mas,N_\zeta)$, depending on
$\mas$, $\coscon$ \rf{omegadef}, and the operator $N_\zeta$, takes the form
\beq
\label{Fdef} && F(\mas,N_\zeta) = \mas^2 + \coscon \Bigl( \frac{d}{2}
- N_\zeta\Bigr)^2  \,.
\eeq

Requiring E.o.M \rf{sec5006} to be consistent with
$\gamma$-tracelessness constraint \rf{gamtra1n} leads to the Lorentz
constraint,

\be
\label{Lorconst} \Bigl( \bar\alpha D + (d + 2 - 2N_\zeta) \fwt_3
\bar\zeta\Bigr)\psik =0\,,
\ee
where $\bar\alpha D\equiv\bar\alpha^A D^A$. Equations of motion
\rf{sec5006}, Lorentz constraint \rf{Lorconst}, and
$\gamma$-tracelessness constraint \rf{gamtra1n} constitute a complete
set of differential and algebraic constraints to be imposed on the
massive spin 5/2 field in the framework of on-shell gauge invariant
approach.

We now discuss gauge symmetries of E.o.M given in
\rf{sec5006}. To this end we introduce parameters of gauge
transformations $\epsilon^{A\alpha}$, $\epsilon^\alpha$ which are the
respective  Dirac complex-valued tensor-spinor spin $3/2$ and $1/2$
fields of the $so(d-1,1)$ Lorentz algebra, where the
$\epsilon^{A\alpha}$ is $\gamma$-traceless, i.e., we start with a
collection of the tensor-spinor fields
\be \label{epscollect}
\epsilon^{A\alpha}\,,\quad \epsilon^\alpha \,,\qquad \gamma^A
\epsilon^{A }= 0\,. \ee
As before, to simplify our expressions we use a ket-vector of gauge
transformations parameter
\beq
\label{gaugpar1}
&& \epsilonk  \equiv |\epsilon_1\rangle +\zeta |\epsilon_0\rangle\,,
\\[3pt]
&& \label{gaugpar2} |\epsilon_1\rangle \equiv \alpha^A
\epsilon^{A\alpha} |0\rangle\,,\qquad |\epsilon_0 \rangle \equiv \epsilon^\alpha
|0\rangle\,.
\eeq
The ket-vector $\epsilonk$ satisfies the algebraic constraints
\beq
&& \label{gaugpar3} (N_\alpha + N_\zeta  - 1)\epsilonk =0\,,
\\
&& \label{gaugpar4} \gamma\bar\alpha \epsilonk =0\,.
\eeq
The constraint \rf{gaugpar3} tells us that the ket-vector $\epsilonk$
is a degree-1 homogeneous polynomial in the oscillators $\alpha^A$,
$\zeta$, while the constraint \rf{gaugpar4} respects the
$\gamma$-tracelessness of $\epsilon^A$, \rf{epscollect}.

We now can express gauge transformation entirely in terms of $\psik$
and $\epsilonk$,
\beq
&& \label{gaugtrapsi} \delta \psik = (\alpha D + \FF)\epsilonk \,,
\\[3pt]
&& \FF \equiv \zeta \fwt_1 + \gamma\alpha\fwt_2 + \alpha^2 \fwt_3
\bar\zeta \,, \eeq
where the operators $\fwt_1=\fwt_1(N_\zeta)$, $\fwt_2=\fwt_2(N_\zeta)$ do
not depend on the $\gamma$-matrices and $\alpha$-oscillators, and
take the form
\beq
\label{Del1def} && \fwt_1(N_\zeta)  = \Bigl(\frac{d +1  - N_\zeta}{d -
2N_\zeta}\, F(\mas,N_\zeta)\Bigr)^{1/2}\,,
\\[3pt]
\label{Del2def}  && \fwt_2(N_\zeta) =   \frac{d + 2}{(d + 2 - 2N_\zeta)( d  -
2N_\zeta)} \mas \,,
\eeq
and $\fwt_3$, $F$ are defined in \rf{Del3def},\rf{Fdef}.

In order that gauge transformation \rf{gaugtrapsi} respects
$\gamma$-tracelessness constraint \rf{gamtra1n} the gauge
transformation parameter $\epsilonk$ should satisfy E.o.M
\be \label{epsequmot01}
\Bigl( \Dline + m_1^\smone + 2\gamma\alpha \fwt_3(0)
\bar\zeta\Bigr)\epsilonk =0\,,
\ee
where the mass operator $m_1^\smone=m_1^\smone (N_\zeta)$ takes the form
\be
\label{mwt1plussol} m_1^\smone(N_\zeta)  = \frac{d + 2}{d - 2N_\zeta} \mas \,.
\ee
In turn, E.o.M \rf{epsequmot01} and
$\gamma$-tracelessness constraint \rf{gaugpar4} imply
the Lorentz constraint for $\epsilonk$,
\be\label{Lorconeps}
\bigl( \bar\alpha D + d \fwt_3(0) \bar\zeta\bigr)\epsilonk =0\,.
\ee
Equations of motion \rf{epsequmot01}, Lorentz constraint
\rf{Lorconeps}, and $\gamma$-tracelessness constraint \rf{gaugpar4}
constitute a complete set of differential and algebraic constraints to
be imposed on the parameter of gauge transformation $\epsilonk$ in
the framework of on-shell gauge invariant approach.

We expressed our results in terms of the mass parameter $\mas$. Since
there is no commonly accepted definition of mass in $(A)dS$ we relate
our mass parameter $\mas$ with various mass parameters used in the
literature. One of the most-used definitions of mass, which we denote
by $m_\smD$, is obtained from the following expansion of E.o.M:
\be\label{mDdef} (\Dline + m_\smD) |\psi_2 \rangle + \ldots =0\,,\ee
where dots stand for $\gamma\bar\alpha\,\psi_2$ and $|\psi_1\rangle$,
$|\psi_0\rangle$ terms.  Comparing \rf{mDdef} with
\rf{sec5006},\rf{mwt1sol} leads then to the identification
\be\label{kappamD} \mas = m_\smD\,.\ee
Another definition of mass parameter for fermionic fields in
$(A)dS_d$ \cite{Metsaev:2006zy}, denoted by $m$, can be obtained by
requiring that the value of $m=0$ corresponds to the massless fields.
For the case of spin 5/2 field in $(A)dS_d$ the mass parameter $m$ is
related with $m_\smD$ as
\be \label{mDmrelnew} m_\smD  = m + \frac{d}{2}\,\sqrt{- \coscon }
\,.
\ee
For the case of $AdS_d$, substituting
$\coscon = - 1/R^2 $ gives $m_\smD = m +
\frac{d}{2R}$ and it turns out that the values $m>0$ correspond to
massive unitary irreps of the $so(d-1,2)$ algebra
\cite{Metsaev:1995re,Metsaev:2003cu}.

Two-derivative and three-derivative
interaction vertices are constructed out the on-shell massive spin
5/2 field $\psik$ and the on-shell massless spin 2 field in $AdS_d$
background. Two-derivative and three-derivative
vertices that depend
on the massless spin 2 field $h^{AB}$ through the linearized Riemann
tensor constructed out the field $h^{AB}$ automatically
respect the gauge symmetries of the massless spin 2 field $h^{AB}$.
Because the massless spin 2 field is assumed to be on-shell%
\be
(\DD^2 -2 \coscon) h^{AB}=0\,, \qquad \DD^Ah^{AB}=0, \qquad
h^{AA}=0\,, \ee
{\it the linearized Riemann tensor for $h^{AB}$ is equal to the linearized Weyl
tensor} $C^{ABCE}$ for $h^{AB}$. This, the two-derivative and
three-derivative vertices we use in this paper
depend on the massless spin 2 field $h^{AB}$ through
the linearized Weyl tensor. We use the Weyl tensor with conventional symmetry
properties: $C^{ABCE}=-C^{BACE}$,
$C^{ABCE}=C^{CEAB}$. The Weyl tensor $C^{ABCE}$ constructed out
the on-shell massless spin 2 field $h^{AB}$ satisfies the well-known
E.o.M and various differential and algebraic
constraints:
\beq
\label{DDCeq} && \left(\DD^2 -2(d-1) \coscon \right)C^{ABCE}=0\,,
\\
\label{DDCeq2} && \DD^A C^{ABCE}=0\,,
\\
\label{bia01} && \DD^{\{F} C^{AB\}C E}=0\,,
\\
\label{bia02} && C^{\{ABC\}E}=0  \,,
\\
\label{trbia02} && C^{AEEB}=0  \,,
\eeq
where the notation $\{ABC\}$ implies summation over cyclic
permutations of indices $A,B,C$. The Weyl tensor can be
expressed in terms of $h^{AB}$ as
\beq
C^{ABCE} & = & \half (\DD^C \DD^B h^{AE} - \DD^C \DD^A h^{BE} + \DD^E
\DD^A h^{BC} - \DD^E \DD^B h^{AC})
\nonumber\\
& + & \half \rho (\eta^{BC} h^{AE} - \eta^{AC} h^{BE} + \eta^{AE}
h^{BC}  - \eta^{BE} h^{AC} )\,.
\eeq
As we have said, we use only those two-derivative and three-derivative
vertex operators (see \rf{M21def}-\rf{M35cdef}) that
depend on the massless spin 2 field $h^{AB}$ through
the Weyl tensor $C^{ABCE}$. Note that there are other
two-derivative and three-derivative vertex operators
that respect linearized  gauge symmetries but do not allow
representation entirely in terms of $C^{ABCE}$.
For example,
in flat space, there is two-derivative cubic vertex ${\rm i}\LL' = \psibr M' \psik $,
\be \label{PnonCver01}
M' = C^{ABCE} \gamma^A \left( \alpha^B \alpha^C
\bar\alpha^E \bar\zeta
- \zeta \alpha^E \bar\alpha^B \bar\alpha^C \right)
+ \wlin^{ABC} \gamma^B \partial^A (\alpha^C \fwh(0)\bar\zeta
+\zeta \fwh(0)\bar\alpha^C)\,,
\ee
$\fwh(0)\equiv 1-N_\zeta$,
which involves the linearized Lorentz connection
\rf{omegalin01} and respects on-shell gauge symmetries%
\footnote{ The vertex \rf{PnonCver01} describes interaction of massless
spin 5/2 field with massless spin 3/2 field and massless spin 2 field.
In the vertex ${\rm i}\LL' = \psibr M' \psik$ \rf{PnonCver01},
the $C$ term and the $\wlin$ term are separately invariant
w.r.t linearized massless spin 2 field
gauge symmetries. Relative coefficient between the $C$ term and the $\wlin$
term is determined by invariance w.r.t. linearized massless spin 5/2 field or
massless spin 3/2 field gauge symmetries.}. In this paper, we do not exploit
vertices like the one in \rf{PnonCver01}.

\newsection{ Restrictions imposed on cubic vertices
by on-shell gauge invariance}\label{ResSec01}

We now demonstrate how the interaction vertices above-discussed can
be obtained by using massive spin 5/2 on-shell gauge symmetries.
Before we formulate our statement, let us outline general strategy
we use to find the gravitational and higher-derivative
interaction vertices. The strategy is as follows (for details, see Appendix D).

{\bf i}) Firstly, we start with the minimal gravitational vertex and study
restrictions imposed by gauge invariance. At this stage we make
sure the gauge variation of the minimal gravitational vertex of the
massive spin 5/2 field contains the Weyl tensor
which cannot be compensated by variation of the graviton metric
tensor. Thus, the minimal gravitational vertex is not self-consistent.

{\bf ii}) Secondly, we add to the minimal gravitational vertex
all possible vertices involving two derivatives and
consider restrictions imposed by gauge invariance.
At this stage we make sure that it is possible
to construct the gauge invariant gravitational vertex when $d=4$, but
this is not possible when $d>4$.

{\bf iii}) Finally, we add all possible two-derivative and
three-derivative vertices%
\footnote{ We use two-derivative and three-derivative vertices
subject to condition
that they depend on the massless spin 2 field $h^{AB}$
through the linearized Weyl tensor
constructed out $h^{AB}$.}
to the minimal gravitational vertex and
consider restrictions imposed by gauge invariance. Now we make
sure that it is possible to construct the gauge invariant gravitational
vertex for $d>4$ too.

Motivated by observation that in order to construct the gravitational
vertex for the massive spin 5/2 field in $d>4$ one needs to involve
vertices with two and three
derivatives we start from the very beginning with the minimal
gravitational vertex supplemented by all possible vertices with two
and three derivatives and look for general solution to restrictions
imposed by requiring the action to be
gauge invariant. In this way, we find solution
to the gravitational vertex and all possible higher-derivative vertices
involving two and three derivatives.

We now formulate our statement.

Let $d>4$. Consider Lagrangian \rf{Lint} with vertex operators given in
\rf{Mtwo}, \rf{Mthree}. All that is required is to determine the quantities
$m_{a,b}$ and $m_{a,b}^\im$ \rf{M21def}-\rf{M35cdef}.
Requiring the Lagrangian to be invariant with respect to
on-shell gauge transformation given in \rf{gaugtrapsi} we obtain that:

\medskip
\noindent
{\bf a}) The quantities
$m_{2,a}$, $a=1,2,3$, and $m_{3,1}(0)$, $m_{3,3}(0)$ can be
solved in terms of five quantities $m_{3,1}(1)$ $m_{3,2}$,
$m_{3,3}(1)$, $m_{3,4}$, $m_{3,5}$. For the latter 5 quantities,
we obtain 3 equations,
i.e., we have two parametric freedom in the solution for $m_{a,b}$.

\medskip
\noindent
{\bf b}) $m_{2,3}^\im$ is equal to zero, while  $m_{3,3}^\im(0)$ can be
solved in terms of $m_{3,4}^\im$. For the three remaining quantities
$m_{3,3}^\im(1)$,
$m_{3,4}^\im$, $m_{3,5}^\im$ we obtain 5 equations, i.e., we
have over-determined system of equations for $m_{a,b}^\im$.

We now outline proof of our statement.
To this end we evaluate the gauge variation of Lagrangian \rf{Lint}.
The gauge variation allows the
representation (by module of total derivatives)

\beq\label{varLag}
{\rm i} e^{-1}\delta \LL^{int} & = & \psibr V \epsilonk - h.c.\,,
\\[5pt]
\label{varLag01}
&&  V = V_\smfour +  V_\smthree +  V_\smtwo\,,
\eeq
where $V_\smfour$, $V_\smthree$, $V_\smtwo$ stand for the respective
four-, three- and two-derivative operators. They allow the representation
\beq \label{varLag02}
&& V_\smfour = \sum_{a=1}^3 X_{4,a} \VV_{4,a} + \sum_{a=2}^3
X_{4,a}^\im \VV_{4,a}\,,
\\
\label{varLag03} && V_\smthree = \sum_{a=1}^5 (X_{3,a} + X_{3,a}^\im)
\VV_{3,a}\,,
\\
\label{varLag04} && V_\smtwo = \sum_{a=1}^2 (X_{2,a} + X_{2,a}^\im)
\VV_{2,a}\,,
\eeq
where $X_{a,b}$ and $X_{a,b}^\im$ are some linear functions of the
respective quantities $m_{a,b}$ and $m_{a,b}^\im$. Explicit expressions for
$X_{a,b}$ and $X_{a,b}^\im$ may be found in Appendix D
(see \rf{X41def}-\rf{X22cdef01}). The $\VV_{a,b}$ are
base operators constructed out the Weyl tensor, the oscillators,
and the covariant derivatives,
\beq\label{VV41def01}
&& \VV_{4,1} \equiv \DD^F C^{A(BC)E} \gamma^A
\alpha^F\alpha^B\bar\alpha^C D^E\,,
\\[3pt]
&& \VV_{4,2}\equiv \DD^F C^{ABCE} \gamma^{AB} \alpha^F \alpha^C
\bar\zeta D^E\,,
\\[3pt]
&& \VV_{4,3}\equiv \DD^F C^{ABCE} \gamma^{AB} \zeta \alpha^F
\bar\alpha^C D^E\,,
\\[15pt]
&& \hspace{0.7cm}
\VV_{3,1}\equiv C^{ABCE}\gamma^{AB} \alpha^C D^E\,,
\\[3pt]
&& \hspace{0.7cm}
\VV_{3,2} \equiv C^{ABCE} \alpha^B\alpha^C\bar\alpha^A D^E\,,
\\[3pt]
&& \hspace{0.7cm}
\VV_{3,3}\equiv C^{ABCE}\gamma^A \alpha^B \alpha^C \bar\zeta
D^E\,,
\\[3pt]
&& \hspace{0.7cm}
\VV_{3,4}\equiv C^{A(BC)E}\gamma^A \zeta \alpha^B \bar\alpha^C
D^E\,,
\\[3pt]
&& \hspace{0.7cm}
\VV_{3,5} \equiv C^{ABCE}\gamma^{AB} \zeta^2 \bar\alpha^C D^E\,,
\\[15pt]
&& \VV_{2,1} \equiv C^{ABCE} \gamma^A  \alpha^B \alpha^C
\bar\alpha^E\,,
\\[3pt]
\label{VV22def01} && \VV_{2,2}\equiv C^{ABCE}\gamma^{AB} \zeta
\alpha^C \bar\alpha^E\,.
\eeq

Requiring gauge variation \rf{varLag} to vanish gives the equations
$X_{a,b}+X_{a,b}^\im=0$. Since $X_{a,b}$ are real valued, while
$X_{a,b}^\im$ are pure imaginary, these equations amount to two set of equations,
$X_{a,b}=0$,  $X_{a,b}^\im=0$.

We first outline results in analysis of the equations $X_{a,b}=0$.
\\
{\bf i}) Equations $X_{4,a}=0$, $a=1,2,3$, lead to
\be \label{m21repnn}
m_{3,1}(0) = m_{3,2}\,, \qquad m_{3,3}(0) = \frac{1}{4} m_{3,4} \,.
\ee
{\bf ii}) Equations $X_{3,a}=0$, $a=1,2,4$, allow us to solve the quantities
$m_{2,1}(0)$, $m_{2,1}(1)$, $m_{2,2}$, $m_{2,3}$ in terms of
$m_{3,1}(1)$, $ m_{3,2}$, $m_{3,3}(1)$, $m_{3,4}$ and $m_{3,5}$,
\beq
\label{m21rep} && m_{2,1}(0) = - \frac{1}{2d} \mas m_{3,2} -
\frac{1}{4}\fwt_1(0) m_{3,4}\,,
\\[5pt]
\label{m211rep}  && m_{2,1}(1) = - \frac{d+2}{2d(d-2)}\mas m_{3,1}(1)
- 2 \fwt_1(1)m_{3,3}(1) + \frac{d-4}{4d}\fwt_1(0) m_{3,4}\,,
\\[5pt]
\label{m22rep}  && m_{2,2} = 4m_{2,1}(0)\,,
\\[5pt]
\label{m23rep}  && m_{2,3} = -\fwt_1(0) m_{3,1}(1) + \frac{d+1}{d}
\fwt_1(0) m_{3,2} + \mas m_{3,4}\,.
\eeq
{\bf iii}) The remaining 4 equations $X_{3,3}=X_{3,5}=0$,
$X_{2,1}=X_{2,2}=0$, are not all independent. This is to say that these 4
equations lead to 3 equations for 5 unknown quantities
$m_{3,1}(1)$, $ m_{3,2}$, $m_{3,3}(1)$, $m_{3,4}$ and $m_{3,5}$,
\beq
\label{m3eq1}  && -\fwt_1(0) m_{3,1}(1) + \frac{d-2}{d}\fwt_1(0)
m_{3,2} + \frac{4\mas}{d(d-2)}m_{3,4} + 2 \fwt_1(1) m_{3,5}= 0\,,
\\[10pt]
\label{m3eq2}&& \fwt_1(0) m_{3,3}(1) - \frac{d-4}{4(d-2)}\fwt_1(1)
m_{3,4} + \frac{2(d-1)}{d(d-2)} \mas m_{3,5}= 0\,,
\\[10pt]
\label{m3eq3} && -\fwt_1^2(0)m_{3,1}(1) + \Bigl( \frac{d^2 + 5d
-2}{d^2}\mas^2 + \frac{d^2 -3d + 6 }{4} \coscon \Bigr) m_{3,2}
\nonumber\\[5pt]
&& \hspace{5cm} + \frac{2(d-1)}{d}\mas\fwt_1(0) m_{3,4} = -
\sqrt{2}\,\kappa\,.
\eeq
When $d>4$, relations \rf{m21repnn}-\rf{m23rep} and equations
\rf{m3eq1}-\rf{m3eq3} constitute a complete system of equations
for $m_{a,b}$ that can be obtained from the requirement of on-shell gauge
invariance of the Lagrangian. In \rf{m3eq3} we restore dependence on
the gravitational constant $\kappa$ (which throughout this paper is
set $\kappa=1/\sqrt{2}$). The contribution related to $\kappa$ is
obtained from the gauge variation of the minimal gravitational vertex.
Solutions to the equations for the gravitational vertices are given for value
$\kappa=1/\sqrt{2}$, while solutions to the equations for
higher-derivative vertices are obtained for $\kappa=0$%
\footnote{ As is well known, general solution of non-homogenous linear
equations (like the ones in \rf{m3eq1}-\rf{m3eq3}) is obtained by adding a
particular solution (which is realized as gravitational vertex here)
of non-homogeneous equations, $\kappa\ne 0$,
to the general solution (which is realized as
higher-derivative vertex here) of homogenous
equations, $\kappa=0$.}.
All solutions of Eqs.\rf{m3eq1}-\rf{m3eq3} were discussed in Sections
\ref{subsubgravvermass01nn}-\ref{secuniver01} and \ref{speintver01}.

\bigskip

We now discuss results in analysis of the equations $X_{a,b}^\im=0$.
\\
{\bf i}) Equations $X_{4,a}^\im=0$, $a=2,3$, lead to
\be\label{m3eq1cnn}
m_{3,3}^\im(0) = \frac{1}{4} m_{3,4}^\im \,.
\ee
{\bf ii}) The remaining equations $X_{3,a}^\im=0$, $a=1,\ldots,5$, and
$X_{2,1}^\im=X_{2,2}^\im=0$ amount to equations
\beq \label{m3eq1c01}
&& m_{2,3}^\im =  0\,,
\\[3pt]
\label{m3eq1c} && \mas m_{3,4}^\im =0 \,,
\\[3pt]
&& \coscon m_{3,4}^\im= 0\,,
\\[3pt]
\label{m331def01}
&& \fwt_1(1) m_{3,3}^\im(1) = 0 \,,
\\[3pt]
&& \fwt_1(1) m_{3,5}^\im =0 \,,
\\[3pt]
\label{m3eq5c} && \fwt_1(0) m_{3,3}^\im(1) + \frac{2(d-1)}{d(d-2)} \mas
m_{3,5}^\im = 0 \,.
\eeq
Equations \rf{m3eq1cnn}-\rf{m3eq5c} constitute a complete system of
equations for $m_{a,b}^\im$ that are obtained  from the requirement of
on-shell gauge invariance of the Lagrangian. All solutions of
Eqs.\rf{m3eq1cnn}-\rf{m3eq5c} were discussed in Sections
\ref{subsubgravvermass01nn}-\ref{secuniver01} and \ref{speintver01}.

Now let us consider the case $d=4$. When $d=4$, one has taken
into account that vertex operators $M_{a,b}$, $M_{a,b}^\im$
\rf{M21def} -\rf{M35cdef}
and operators
$\VV_{a,b}$ \rf{VV41def01}-\rf{VV22def01} are not all independent.
This is to say that
we proceed as follows.
\\
{\bf a})  The operators $M_{3,1}$, $M_{3,2}$, $M_{3,4}$, $M_{3,4}^\im$
are expressible in terms of the remaining operators $M_{a,b}$, $M_{a,b}^\im$.
In order to take this into account we set the corresponding
coefficients equal to zero,
\be
\label{mabzeroval01}
m_{3,1}(0)=0, \quad m_{3,1}(1)=0, \quad m_{3,2}=0,\quad m_{3,4}=0, \quad
m_{3,4}^\im=0\,.\ee

\noindent
{\bf b}) The operators $\VV_{3,2}$ and $\VV_{3,4}$ are expressible in terms of
the operators $\VV_{3,a}$, $\VV_{2,a}$.
This, when $d=4$, basis of operators $\VV_{a,b}$ is given by
\be \label{veropebasd4nn}
\VV_{4,a}\,,\quad a=1,2,3\,,\qquad \VV_{3,1}\,,\quad \VV_{3,3}\,,\quad
\VV_{3,5}\,,\quad \VV_{2,1}\,, \quad \VV_{2,2}\,. \ee
To take this into account, we substitute the representation for
the operators $\VV_{3,2}$ and $\VV_{3,4}$ in terms of the independent operators
$\VV_{a,b}$
(see \rf{V32d4},\rf{V34d4})
into expressions for the gauge variation (see \rf{varLag02}-\rf{varLag04}).
This gives representation of the gauge variation in terms of the independent
operators \rf{veropebasd4nn}. Then, the gauge variation takes
form \rf{varLag},\rf{varLag01}, where
\beq \label{varLag02d4}
&& V_\smfour = \sum_{a=1,2,3} X_{4,a} \VV_{4,a} + \sum_{a=2,3}
X_{4,a}^\im \VV_{4,a}\,,
\\
\label{varLag03d4} && V_\smthree = \sum_{a=1,3,5} (X_{3,a} + X_{3,a}^\im)
\VV_{3,a}\,,
\\
\label{varLag04d4} && V_\smtwo = \sum_{a=1,2}  (X_{2,a} + X_{2,a}^\im)
\VV_{2,a}\,,
\eeq
and the explicit expressions for $X_{a,b}$ and $X_{a,b}^\im$
are given in Appendix D
(see \rf{X31def01nnn}-\rf{X22d401}).
The respective equations imposed by gauge invariance
take then the form
\beq \label{d4equX01}
&&  X_{4,a}=0\,,\hspace{2.5cm} a=1,2,3\,; \qquad  X_{4,a}^\im=0\,,\qquad a=2,3\,;
\\
&& \label{d4equX02}
X_{3,a} =  X_{3,a}^\im =0\,,\qquad
\quad a=1,3,5\,;
\\
&& \label{d4equX03}
X_{2,a}= X_{2,a}^\im =0\,,\quad
\qquad a=1,2\,.
\eeq
We now discuss results in analysis of equations \rf{d4equX01}-\rf{d4equX03}.
We begin with analysis of the equations $X_{a,b}=0$.
\\
{\bf i})  $X_{4,1}$ is equal to zero automatically, while the
equations $X_{4,2}=X_{4,3}=0$ give
\be \label{m33dold401} m_{3,3}(0) = 0\,.\ee

\noindent
{\bf ii}) Equations $X_{3,1}=0$, $X_{3,3}=0$ allow us to solve the quantities
$m_{2,1}(0)$, $m_{2,1}(1)$, $m_{2,3}$ in terms of
$m_{2,2}$, $m_{3,3}(1)$, and $m_{3,5}$,
\beq
&& \label{Pm21m22d4} m_{2,1}(0) = \frac{1}{4} m_{2,2}\,,
\\
&& m_{2,1}(1) = - 2\fwt_1(1) m_{3,3}(1)\,,
\\
\label{m23m35d4}
&& m_{2,3} = - 2 \fwt_1(1) m_{3,5}\,.
\eeq
We note that, the coefficients $m_{2,1}(0)$, $m_{2,2}$ \rf{Pm21m22d4},
in contrast to the ones in $d>4$ (see \rf{m21rep},\rf{m22rep}),
are not expressible in terms of the coefficients $m_{3,a}$.

\noindent
{\bf iii}) The remaining 3 equations, $X_{3,5}=0$,
$X_{2,1}=X_{2,2}=0$, are not all independent. This is to say that these 3
equations lead to 2 equations for 3 unknown quantities
$m_{2,2}$, $m_{3,3}(1)$, and $m_{3,5}$,
\beq
&& \fwt_1(0) m_{3,3}(1) + \frac{3}{4} \mas m_{3,5} = 0 \,,
\\
&& - \mas m_{2,2} - 2 \fwt_1(0)\fwt_1(1) m_{3,5} + \sqrt{2} \kappa
=0\,.
\eeq

We now proceed to discussion of the equations $X_{a,b}^\im=0$ in
\rf{d4equX01}-\rf{d4equX03}.
\\
{\bf i}) Equations $X_{4,2}^\im = X_{4,3}^\im=0$ lead to
\be \label{mim33dold401} m_{3,3}^\im(0) = 0\,.\ee
{\bf ii}) Equations $X_{3,a}^\im=0$, $a=1,3,5$ and $X_{2,a}^\im=0$, $a=1,2$
amount to
\beq
\label{mabcom01}
&& m_{2,3}^\im = 0\,,
\\
&& \fwt_1(1) m_{3,3}^\im(1) = 0\,,
\\
&& \fwt_1(1) m_{3,5}^\im = 0\,,
\\
\label{mabcom05}
&& \fwt_1(0) m_{3,3}^\im(1) + \frac{3}{4} \mas m_{3,5}^\im =0\,.
\eeq
We note that Eqs.\rf{mabcom01}-\rf{mabcom05}
are obtained from their counterparts for $d>4$ (see
\rf{m3eq1c01},\rf{m331def01}-\rf{m3eq5c}) by a simple substitution $d=4$.

When $d=4$, equations \rf{mabzeroval01},\rf{m33dold401}-\rf{mabcom05}
constitute a complete system of equations
that are obtainable from the on-shell gauge
invariance of the Lagrangian.
All solutions of these equations were discussed in Sections
\ref{secmasverd401},\ref{speintver01}.

\bigskip
\bigskip
{\bf Acknowledgments}. This work was supported by the INTAS project
03-51-6346, by the RFBR Grant No.05-02-17217, RFBR Grant for Leading
Scientific Schools, Grant No. LSS-4401.2006.2 and Russian Science
Support Foundation.

\setcounter{section}{0} \setcounter{subsection}{0}
\appendix{ Notation and conventions}

We use space-time base manifold
indices $\mu,\nu,\sigma,\lambda = 0,1,\ldots ,d-1$ and
flat vectors indices of the $so(d-1,1)$ algebra
$A,B,C,E,D,F = 0,1,\ldots ,d-1$.

We use $2^{[d/2]}\times 2^{[d/2]}$ Dirac gamma matrices $\gamma^A$ in
$d$-dimensions,
\be
\label{gammadef} \{ \gamma^A,\gamma^B\} = 2\eta^{AB}\,,\qquad
\gamma^{A\dagger} = \gamma^0 \gamma^A \gamma^0,
\ee
where $\eta^{AB}$ is mostly positive flat metric tensor.
To simplify our expressions, we drop
$\eta_{AB}$ in scalar products and use $X^AY^A \equiv
\eta_{AB}X^A Y^B$.
Second relation in \rf{gammadef} implies
\beq
({\rm i}\gamma^0)^\dagger  = + {\rm i}\gamma^0\,,
\qquad
({\rm i}\gamma^0\gamma^A)^\dagger  =   - {\rm i}\gamma^0\gamma^A\,,
\qquad
%
({\rm i}\gamma^0\gamma^{AB})^\dagger  =  - {\rm
i}\gamma^0\gamma^{AB}\,.
\eeq

We assume the normalization of spin 5/2 field Lagrangian and
Einstein-Hilbert Lagrangian as
\beq
&& \LL_{5/2} = -{\rm i} E \psibr \Dline \psik + \ldots\,,
\\
&&
\LL_{EH}=
\frac{1}{2\kappa^2}\sqrt{-G} \left( R - (d-1)(d-2)\rho \right)\,,  \eeq
where $E = \det E_\mu^A$, $G= \det G_{\mu\nu}$ and
$\psibr=(\psik)^\dagger\gamma^0$.
Hermitian conjugation rule for a product of fermionic fields $\psi$, $\chi$
is assumed to be
$(\psi\chi)^\dagger = \chi^\dagger \psi^\dagger$.
Riemann tensor and its
contractions are
\be
R^\mu{}_{\nu\lambda\sigma} = \partial_\lambda\Gamma_{\nu\sigma}^\mu+\ldots\,,
\qquad
R_{\mu\nu} = R^\sigma{}_{\mu\sigma\nu} \,,
\qquad
R = G^{\mu\nu}R_{\mu\nu}\,.
\ee

Metric $G_{\mu\nu}$ and
veilbein $E_\mu^A$ are expanded over $(A)dS$ space metric $g_{\mu\nu}$
and vielbein $e_\mu^A$ as

\be
E_\mu^A  = e_\mu^A  + \frac{\kappa}{\sqrt{2}}h_\mu^A\,,
\qquad
G_{\mu\nu} = g_{\mu\nu} + \sqrt{2}\,\kappa
h_{\mu\nu} + \ldots\,,\qquad h_{\mu\nu} \equiv e_\mu^A h_\nu^A\,.
\ee
The gravitational constant $\kappa$,
unless otherwise specified, is set to be $\kappa=1/\sqrt{2}$.

$(A)dS$ space-time tensor, $h^{\mu\nu}$, and tensor-spinor, $\psi^{\mu\nu}$,
fields are related with
the respective fields carrying the flat indices in a standard way,
\be h^{AB} = e_\mu^A e_\nu^B h^{\mu\nu}\,,
\qquad\quad
\psi^{AB} = e_\mu^A e_\nu^B \psi^{\mu\nu}\,. \ee
Using the flat indices and $\coscon$ given in \rf{omegadef}, we obtain
Riemann tensor for $(A)dS$ space
\be R^{ABCD}|_{(A)dS_d}^{\vphantom{5pt}} = \coscon
(\eta^{AC}\eta^{BD} - \eta^{AD}\eta^{BC})\,. \ee

Covariant derivative $\DD_\mu$, acting on vector and spinor fields
\be \DD_\mu \phi^B = \partial_\mu \phi^A + \omega_\mu^{AB}(e)\phi^B\,, \quad\qquad
\DD_\mu \psi = \partial_\mu \psi + \frac{1}{4}\omega_\mu^{AB}(e)\gamma^{AB}\psi\,,  \ee
satisfies the standard commutators
\be [\DD_\mu,\DD_\nu] \phi^B = R_{\mu\nu}^{AB}\phi^B \,,\qquad \quad
[\DD_\mu,\DD_\nu] \psi = \frac{1}{4}R_{\mu\nu}^{AB} \gamma^{AB} \psi\,.  \ee
Instead of $\DD_\mu$, we
prefer to use a covariant derivative with the flat indices $\DD^A$,
\be
\DD_A \equiv e_A^\mu \DD_\mu\,,\qquad \DD^A = \eta^{AB}\DD_B\,,\ee
where $e_A^\mu$ is inverse of $(A)dS$ vielbein, $e_\mu^A e_B^\mu = \delta_B^A$.

\appendix{ Commutators of oscillators and
derivatives }

We use the algebra of commutators for operators that can be
constructed out the oscillators $\alpha^A$, $\bar\alpha^A$, $\gamma$-matrices,
and
derivative $D^A$ \rf{vardef01},\rf{lorspiope}
(see also Appendix A in Ref.\cite{Metsaev:1999ui}).
Starting with
\be\label{app00m01}
[\hat{\partial}_A,\hat{\partial}_B]=\Omega_{AB}{}^C\hat{\partial}_C\,,
\qquad \Omega^{ABC}\equiv -\omega^{ABC}+\omega^{BAC}\,, \qquad
\omega_A{}^{BC}\equiv E^\mu_A\omega_\mu^{BC} \,, \ee
where $\hat\partial_A\equiv E_A^\mu\partial_\mu$,
$\partial_\mu\equiv\partial/\partial x^\mu$,  and $\Omega^{ABC}$ is a
contorsion tensor, we get the basic commutator
\beq \label{dadbap2}
[D^A , D^B ]  & = & \Omega^{ABC} D^C + \frac{1}{2} R^{ABCD} M^{CD}
\nonumber\\[3pt]
& = & \Omega^{ABC} D^C + \coscon M^{AB}\,.
\eeq
The spin operator
$M^{AB}$ is given in \rf{loralgspiope}. In
\rf{dadbap2} and below, we represent our relations for a space of
arbitrary geometry and for $(A)dS_d$ space. Using \rf{dadbap2} and
the commutators
\be\label{app001}
[D^A,\alpha^B]= -\omega^{ABC}\alpha^C\,, \qquad [D^A,
\bar{\alpha}^B]= - \omega^{ABC}\bar{\alpha}^C\,, \qquad [D^A,
\gamma^B]= - \omega^{ABC}\gamma^C\,, \ee
we find straightforwardly
\beq
&& [D^A, \alpha^2]=0\,, \qquad  [\bar\alpha^2, D^A]=0\,,\qquad
[D^A,\gamma\alpha]=0\,,
\\[3pt]
\label{app003} && [\bar\alpha^2, \alpha D] = 2\bar\alpha D  \,,
\qquad
[\gamma\bar\alpha, \alpha D] = \Dline  \,,
\qquad
\{\Dline, \gamma\alpha\} = 2\alpha D \,,
\eeq

\beq \label{app004}
\Dline\!^2 & = & \Box +
\frac{1}{4}\gamma^{AB}R^{ABCD} M^{CD}
\nonumber\\[3pt]
& = & \Box + \frac{ \coscon
}{2}\gamma^{AB}M^{AB}\,,
\\[3pt]
{} [\bar\alpha D , \alpha D ]   & = & \Box  -
\frac{1}{4} R^{ABCD} M_b^{AB} M^{CD}
\nonumber\\
&  = & \Box - \frac{\coscon}{2} M_b^{AB}
M^{AB}\,,
\\[3pt]
[\Dline,\alpha D ]  & = & \frac{1}{2}\gamma^A \alpha^B R^{ABCD}M^{CD}
\nonumber\\
& = & \coscon \Bigl(  \gamma \alpha (\alpha\bar\alpha +
\frac{d-1}{2}) - \alpha^2\gamma \bar\alpha\Bigr)\,,
\\[3pt]
{} [\bar\alpha D, \Dline ] & = & - \frac{1}{2}\gamma^A \bar\alpha^B
R^{ABCD} M^{CD}
\nonumber\\
& =  & \coscon \Bigl( (\alpha\bar\alpha + \frac{d-1}{2}) \gamma
\bar\alpha  -  \gamma\alpha\bar\alpha^2 \Bigr) \,,
\\
{} [D^A,\alpha D] & = &  -\omega^{BCA} \alpha^B D^C +
\frac{1}{2}R^{ABCD}\alpha^B M^{CD}
\nonumber\\
& = &  - \omega^{BCA} \alpha^B D^C +  \coscon  \Bigl(\alpha^A
(\alpha\bar\alpha + \frac{1}{2}) -\alpha^2 \bar\alpha^A - \frac{1}{2}
\gamma \alpha \gamma^A\Bigr) \,,
\\[7pt]
{} [D^A,\bar\alpha D] & = &  -\omega^{BCA} \bar\alpha^B D^C +
\frac{1}{2}R^{ABCD}\bar\alpha^B M^{CD}
\nonumber\\
& = &  - \omega^{BCA} \bar\alpha^B D^C +  \coscon  \Bigl(-
(\alpha\bar\alpha + d- \half ) \bar\alpha^A
 + \alpha^A\bar\alpha^2 +  \frac{1}{2}
\gamma^A \gamma\bar\alpha \Bigr) \,,
\\[3pt]
\label{app0010}
[D^A,\Dline] & = &  -\omega^{BCA} \gamma^B D^C +
\frac{1}{2}R^{ABCD}\gamma^B M^{CD}
\nonumber\\
& = &  - \omega^{BCA} \gamma^B D^C + \coscon  \Bigl( \alpha^A
\gamma\bar\alpha  - \gamma\alpha \bar\alpha^A - \frac{d-1}{2}
\gamma^A\Bigr) \,,
\eeq
where we use the notation
\be \label{Dcovdef01} \Box   \equiv D^AD^A + \omega^{AAB}D^B\,.\ee
Expressions \rf{app00m01}, \rf{app001}-\rf{app003} and the ones in
1st lines in \rf{dadbap2},\rf{app004}-\rf{app0010} are valid for
arbitrary geometry, while the
expressions in the 2nd lines in \rf{dadbap2},
\rf{app004}-\rf{app0010} are adopted to $(A)dS_d$ geometry.

\appendix{ On-shell gauge invariant formulation for massive spin
$s+\half$ fermionic field in $(A)dS_d$}

The aim of this Appendix is twofold. Firstly, we outline
on-shell gauge invariant formulation for massive arbitrary spin
fermionic field. Secondly, we present some details of derivation of
on-shell gauge invariant formulation for the massive spin 5/2
field.

{\it On-shell gauge invariant formulation} can be obtained from the
{\it gauge invariant formulation} developed in \cite{Metsaev:2006zy}.
We start therefore with brief discussion of the {\it gauge invariant
formulation}.

Gauge invariant Lagrangian for massive arbitrary spin $s+\half$
fermionic field ($s$ is an integer number) is constructed out tensor-spinor spin
$s'+\frac{1}{2}$ fields of the $so(d-1,1)$ Lorentz algebra
$\psi^{A_1\ldots A_{s'}\alpha}$, $s'=0,1,\ldots, s$. Collecting these
tensor-spinor fields into a ket-vector $\psik$ defined by
\beq
&& |\psi\rangle  \equiv \sum_{s'=0}^s
\zeta^{s-s'}|\psi_{s'}\rangle\,,
\\
&&  |\psi_{s'}\rangle \equiv \alpha^{A_1}\ldots \alpha^{A_{s'}}
\psi^{A_1\ldots A_{s'}\alpha}(x)|0\rangle\,,
\eeq
allows us to write the Lagrangian entirely in terms of $\psik$. The
tensor-spinor fields satisfy, by construction, $\gamma$-triple
tracelessness constraint which can be written in terms
of $\psik$ as
\beq
&&  \gamma\bar\alpha \bar\alpha^2 |\psi\rangle =0\,.
\eeq
Lagrangian for the massive fermionic field in $(A)dS_d$ space found
in \cite{Metsaev:2006zy} takes the form
\be \label{appeb10} \LL = \LL_{der} + \LL_m\,, \ee
\be
\label{appeb11} {\rm i}e^{-1} \LL_{der} =   \psibr L \psik\,,
\qquad\quad   {\rm i}e^{-1} \LL_m  =  \psibr \MM \psik\,, \ee
where $L$ stands for a first order differential operator,
while $ \MM$ stands for a mass operator.
The operators $L$ and $\MM$ take the form
\beq \label{appeb12}
L  & = & \Dline  - \alpha D \gamma\bar\alpha -
\gamma\alpha\,\bar\alpha D + \gamma\alpha \Dline\gamma\bar\alpha
\nonumber\\
& + & \frac{1}{2}\gamma\alpha\,\alpha D \bar\alpha^2 +
\frac{1}{2}\alpha^2\gamma\bar\alpha\,\bar\alpha D -
\frac{1}{4}\alpha^2\Dline\bar\alpha^2\,,
\\
\label{appeb13} \MM &= & (1 - \gamma\alpha \gamma\bar\alpha
-\frac{1}{4} \alpha^2 \bar\alpha^2 ) m_1 + \mwt_4 (\gamma\alpha
\bar\zeta - \frac{1}{2} \alpha^2 \bar\zeta \gamma\bar\alpha)
\nonumber\\
& - & (\zeta  \gamma\bar\alpha  - \frac{1}{2}\gamma\alpha \zeta
\bar\alpha^2)\mwt_4\,, \eeq
where we use the notation
\beq
\label{appeb14} && m_1  = \frac{2s+ d -2}{2s + d - 2 - 2N_\zeta}
\mas \,,
\\
\label{appeb15} && \mwt_4  = \Bigl(\frac{2s+ d -3 - N_\zeta}{2s + d -
4 - 2N_\zeta}\, F(\mas,s,N_\zeta)\Bigr)^{1/2} \,,
\\
\label{appeb16} && F(\mas,s,N_\zeta) = \mas^2 + \coscon \Bigl(s +
\frac{d-4}{2} - N_\zeta\Bigr)^2  \,.
\eeq
To describe a gauge transformation, we introduce
tensor-spinor fields $\epsilon^{{A_1} \ldots A_{s'}}$,
$s'=0,1,\ldots,s-1$. As before, we collect these tensor-spinor fields
in a ket-vector $\epsilonk$ defined by
\beq
\label{appeb17} && \epsilonk  \equiv \sum_{s'=0}^{s-1}
\zeta^{s-1-s'}|\epsilon_{s'}\rangle\,,
\\
\label{appeb18} &&  |\epsilon_{s'}\rangle \equiv \alpha^{A_1}\ldots
\alpha^{A_{s'}} \epsilon^{A_1\ldots A_{s'}\alpha}(x)|0\rangle\,.
\eeq
The tensor-spinor fields $\epsilon^{{A_1} \ldots A_{s'}}$ satisfy, by
construction, $\gamma$-tracelessness constraint which can be written
in terms of the ket-vector $\epsilonk$ as $\gamma\bar\alpha\epsilonk=0$.
With this notation, the gauge
transformation takes the form
\beq
\label{appeb19} \delta \psik & = & (\alpha D + \FF )\epsilonk \,,
\\
\label{appeb20} && \FF \equiv \zeta \fwt_1 + \gamma\alpha \fwt_2 +
\alpha^2 \fwt_3 \bar\zeta \,,
\\[3pt]
\label{appeb21} && \fwt_1  \equiv \Bigl(\frac{2s+ d -3 - N_\zeta}{2s
+ d - 4 - 2N_\zeta}\, F(\mas,s,N_\zeta)\Bigr)^{1/2}\,,
\\
\label{appeb22} && \fwt_2  \equiv   \frac{2s+ d -2}{(2s + d - 2 -
2N_\zeta)(2s + d - 4 - 2N_\zeta)} \mas \,,
\\
\label{appeb23} && \fwt_3  \equiv  -\Bigl(\frac{2s+ d -3 - N_\zeta}{
(2s + d - 4 - 2N_\zeta)^3}\, F(\mas,s,N_\zeta)\Bigr)^{1/2}\,.
\eeq
We now proceed with discussion of {\it on-shell gauge invariant
formulation}. In the framework of on-shell gauge invariant
formulation, the massive arbitrary spin $s+\half$ fermionic field $\psik$
satisfies the following differential and algebraic equations:
\beq
\label{appeb30} && \Bigl( \Dline + m_1 + 2\gamma\alpha \fwt_3
\bar\zeta\Bigr)\psik =0\,,
\\
\label{appeb31} && \Bigl( \bar\alpha D + (2s+d-2-2N_\zeta) \fwt_3
\bar\zeta\Bigr)\psik =0\,,
\\
\label{appeb32} && \gamma\bar\alpha\psik=0\,, \eeq
which are the respective E.o.M \rf{appeb30}, Lorentz
constraint \rf{appeb31} and $\gamma$-tracelessness constraint
\rf{appeb32}. In on-shell gauge invariant approach, the parameter of
gauge transformation $\epsilonk$ satisfies the differential and
algebraic equations,
\beq
\label{appeb33} && \Bigl( \Dline + m_1^\smone + 2\gamma\alpha \fwt_3
\bar\zeta\Bigr)\epsilonk =0\,,
\\
\label{appeb34} && \Bigl( \bar\alpha D + (2s+d-4-2N_\zeta)  \fwt_3
\bar\zeta\Bigr)\epsilonk =0\,,
\\
\label{appeb35} && \gamma\bar\alpha\epsilonk=0\,, \eeq
where
\be
m_1^\smone   = \frac{2s+ d -2}{2s + d - 4- 2N_\zeta}
\mas \,.
\ee
Equations \rf{appeb30}-\rf{appeb35} constitute a complete set of
equations for $\psik$ and $\epsilonk$ in the framework of on-shell
gauge invariant approach. These equations can obviously be derived
from the gauge invariant approach discussed earlier in
this Appendix. To demonstrate how this derivation comes about, we
consider the case of spin 5/2 field in what follows.

In view of gauge symmetry \rf{appeb19} and the constraint
$\gamma\bar\alpha\epsilonk=0$ we can impose the following gauge condition on the
massive spin 5/2 field $\psik$:
\be \label{appeb36} (\gamma\bar\alpha - \frac{1}{d}\gamma\alpha
\bar\alpha^2) \psik= 0\,. \ee
Requiring gauge condition \rf{appeb36} to be invariant w.r.t gauge
transformation \rf{appeb19},
\be \label{appeb37}  (\gamma\bar\alpha - \frac{1}{d}\gamma\alpha
\bar\alpha^2) \delta \psik= 0\,, \ee
leads to the following equations for the parameter of gauge
transformation:
\be \label{appeb38}  (\Dline - \frac{2}{d} \gamma\alpha \bar\alpha D
+ \frac{d+2}{d-2N_\zeta}\mas )\epsilonk =0\,. \ee
Before we proceed, observe that Eqs.\rf{appeb38} lead to the
following equations:
\be
\label{appeb39} {\bf D}\bar\zeta\epsilonk =0 \,,
\qquad\qquad  {\bf D}\bar\alpha D \epsilonk =0\,,
\ee
\be
\label{appeb40} {\bf D} \equiv  \Dline + \frac{d+2}{d-2}\mas\,. \qquad \quad
\ee
To summarize, there is leftover gauge symmetry \rf{appeb19} with
the parameter $\epsilonk$ that satisfies Eqs.\rf{appeb38}. We now
consider leftover gauge variation of $\bar\alpha^2\psik$
and E.o.M for $\bar\alpha^2\psik$. Using
\rf{appeb19} we obtain
\be
\label{appeb41} \delta \left(\bar\alpha^2 \psik\right) = |\EE\rangle
\,,
\qquad
|\EE\rangle \equiv 2\Bigl(\bar\alpha D + d \fwt_3(0) \bar\zeta
\Bigr)\epsilonk\,,
\ee
while using \rf{appeb36} and E.o.M derived from
Lagrangian \rf{appeb10}, $(L+\MM)\psik=0$, we obtain
\be \label{appeb42} {\bf D} \bar\alpha^2 \psik =0\,.
\ee
Also, it is easy to see that Eqs.\rf{appeb39} lead to the following
equations for $|\EE\rangle$:
\be \label{appeb43} {\bf D} |\EE\rangle =0\,.
\ee
From \rf{appeb42},\rf{appeb43}, we see that $\bar\alpha^2\psik$ and
$|\EE\rangle$ satisfy the same E.o.M. Taking this and the 1st relation
in \rf{appeb41} into account, we can set
\be \label{appeb44} \bar\alpha^2\psik= 0\,,  \qquad |\EE\rangle =0
\,.\ee
Gauge condition \rf{appeb36} and the 1st relation in \rf{appeb44} implies
$\gamma$-tracelessness constraint for $\psik$, \rf{gamtra1n}, while
the 2nd relations in \rf{appeb41} and \rf{appeb44} imply Lorentz
constraint for $\epsilonk$, \rf{Lorconeps}.

\appendix{ Analysis of gauge variation of Lagrangian }

Taking into account
gauge transformation \rf{gaugtrapsi}, we see that
the gauge variation of Lagrangian \rf{Lint}  can be written as
\beq \label{appCC01}
&& \delta \LL^{int} = \delta \LL_\smone + \delta \LL_\smtwo + \delta
\LL_\smthree \,,
\\[5pt]
&& \label{appCC02}
{\rm i} e^{-1}\delta \LL_\sma = \psibr M_\sma (\alpha D + \FF)
\epsilonk - h.c.\,, \qquad a=1,2,3.
\eeq
Using E.o.M, the Lorentz constraint, and the $\gamma$-tracelessness
constraints for the ket-vectors $\psik$ and $\epsilonk$, gauge variation
\rf{appCC01} can be represented as in \rf{varLag}-\rf{varLag04},
where the operators $\VV_{a,b}$ given in
\rf{VV41def01}-\rf{VV22def01} constitute a complete basis of the operators
involving four, three, and two derivatives.  The quantities $X_{a,b}$
and $X_{a,b}^\im$
depend on $m_{a,b}$  and  $m_{a,b}^\im$  respectively. Some of
$X_{a,b}$,  $X_{a,b}^\im$ depend on the $N_\zeta$.
When $d>4$, we obtain the following expressions for
$X_{a,b}$,  $X_{a,b}^\im$:
\beq \label{X41def}
&& X_{4,1} = -m_{3,1}(0) + m_{3,2}\,,
\\
\label{X4223def}
&&  X_{4,2} = -m_{3,3}(0) + \frac{1}{4} m_{3,4}\,,
\qquad  X_{4,3} = X_{4,2} \,,
\eeq
\beq \label{X31def01}
X_{3,1} &  = &  \frac{1}{2} \fwt_2 m_{3,1} +(d+4)
\fwt_3(0)N_\zeta m_{3,3}(0) - 2\fwt_3(0)N_\zeta m_{3,4}
\nonumber\\
&&
+ (N_\zeta+1)\fwt_1 m_{3,3}  + m_{2,1}\,,
\\[10pt]
X_{3,2} & =  & 3\fwt_2(0)  m_{3,2} - \fwt_2(0) m_{3,1}(0)
+ \fwt_1(0) m_{3,4} - 4m_{2,1}(0) + 2 m_{2,2}\,,
\\[10pt]
X_{3,3} &  =  & \fwt_3(0) m_{3,1}(0) + 2\fwt_3(0)m_{3,2} -
\frac{1}{2} (m_1(0) + m_1(1))m_{3,4}
\nonumber\\
&& + 4\fwt_2(1) m_{3,3}(0)+ 2\fwt_1(1) m_{3,5} +
m_{2,3}\,,
\\[10pt]
X_{3,4} & =  & (d+3)\fwt_3(0) m_{3,1}(0)- 2\fwt_3(0) m_{3,2} -
\half ( m_1(0) +m_1(1)) m_{3,4}
\nonumber\\[5pt]
&& + \fwt_1(0) m_{3,1}(1) + 4 \fwt_2(0) m_{3,3}(0)  + m_{2,3}\,,
\\[10pt]
X_{3,5} &  =  & d\fwt_3(1) m_{3,3}(0) - \fwt_3(1) m_{3,4}
+\frac{1}{2}\fwt_2(1) m_{3,5}
\nonumber\\
&& + \fwt_1(0) m_{3,3}(1) + \frac{3}{2} \fwt_2(0) m_{3,5}\,,
\\[10pt]
X_{2,1} &  = & -\frac{5}{4}\coscon m_{3,1}(0) + \Bigl(\half
(m_1(0) + m_1(1))\fwt_2(0) + \frac{10-5d}{4}\coscon \Bigr) m_{3,2}
\nonumber\\[5pt]
&& -4m_1(0) m_{2,1}(0) + 2 \fwt_2(0) m_{2,2}
+  \fwt_1(0) m_{2,3} + \sqrt{2}\kappa\,,
\\[10pt]
\label{X22def01} X_{2,2} & = &  \frac{3}{2}\fwt_2(0)\fwt_3(0)
m_{3,2} - \frac{3}{2}\coscon m_{3,3}(0) - \frac{d-7}{8}\coscon
m_{3,4}
\nonumber\\
&& + (d-4)\fwt_3(0)m_{2,1}(0) + \fwt_1(0) m_{2,1}(1) + \frac{3}{2}
\fwt_2(0) m_{2,3}\,,
\eeq

\beq \label{X43comdef01}
&& \hspace{1cm} X_{4,2}^\im = -m_{3,3}^\im(0) + \frac{1}{4} m_{3,4}^\im\,, \qquad
X_{4,3}^\im = - X_{4,2}^\im  \,,
\\[5pt]
&&\hspace{-1cm}
X_{3,1}^\im = d \fwt_3(0)N_\zeta m_{3,3}^\im(0) - \fwt_3(0)N_\zeta m_{3,4}^\im
+ (N_\zeta+1)\fwt_1 m_{3,3}^\im \,,
\\[5pt]
&& \hspace{-1cm} X_{3,2}^\im =  \fwt_1(0) m_{3,4}^\im \,,
\\[5pt]
&& \hspace{-1cm} X_{3,3}^\im = m_{2,3}^\im
- \frac{1}{2} (m_1(0) + m_1(1))m_{3,4}^\im
+ 4\fwt_2(1) m_{3,3}^\im(0)+  2\fwt_1(1)
m_{3,5}^\im\,,
\\[5pt]
&&\hspace{-1cm}  X_{3,4}^\im = -m_{2,3}^\im  + \half(m_1(0)+ m_1(1)) m_{3,4}^\im
- 4 \fwt_2(0) m_{3,3}^\im(0)  \,,
\\[5pt]
&& \hspace{-1cm} X_{3,5}^\im
= - d\fwt_3(1) m_{3,3}^\im(0) +  \fwt_3(1) m_{3,4}^\im -
\frac{1}{2}\fwt_2(1) m_{3,5}^\im
- \fwt_1(0) m_{3,3}^\im(1) - \frac{3}{2} \fwt_2(0) m_{3,5}^\im\,,
\\[5pt]
&& \hspace{1cm} X_{2,1}^\im =  \fwt_1(0) m_{2,3}^\im  \,,
\\[5pt]
\label{X22cdef01}
&& \hspace{1cm} X_{2,2}^\im = \frac{3}{2}\rho m_{3,3}^\im(0) +
\frac{d-7}{8}\rho m_{3,4}^\im
-\frac{3}{2} \fwt_2(0) m_{2,3}^\im\,.
\eeq

Requiring gauge variation \rf{varLag} to vanish gives the equations
$X_{a,b}=X_{a,b}^\im=0$. Solutions to these equations were discussed in
Section \ref{ResSec01}. The following remarks are in order.
\\
{\bf i}) From \rf{X41def},\rf{X4223def} and
\rf{X43comdef01}, we see that the equations $X_{4,a}=0$,
$a=1,2,3$, and $X_{4,2}^\im=0$, $X_{4,3}^\im=0$ lead indeed to
\rf{m21repnn} and \rf{m3eq1cnn} respectively.
\\
{\bf ii}) The $X_{3,1}$ defined in \rf{X31def01} depends on the operator
$N_\zeta$, $X_{3,1} = X_{3,1}(N_\zeta)$. Because $X_{3,1}$ is acting
on ket-vector $\epsilonk$ \rf{varLag} which is a degree-1 polynomial in
the oscillator $\zeta$, the equation $X_{3,1}=0$ amounts to the two equations,
$X_{3,1}(0)=0$, $X_{3,1}(1)=0$. These two equations allow us to
solve $m_{2,1}(0)$, $m_{2,1}(1)$ in terms of $m_{3,a}$ (see
\rf{m21rep},\rf{m211rep}). Equation $X_{3,2}=0$ leads to \rf{m22rep} .
\\
{\bf iii}) Using \rf{m21repnn}, we can express $X_{3,4}$ in terms of
$m_{2,3}$ $m_{3,1}(1)$, $m_{3,2}$, $m_{3,4}$. Also, it is convenient
to express $\fwt_3(0)$ \rf{Del3def} in terms of $\fwt_1(0)$
\rf{fwtonezerdef}. After this, it is easy to see that the equation
$X_{3,4}=0$ leads to $m_{2,3}$ given in \rf{m23rep}.

We now outline derivation of
gauge variation \rf{varLag}-\rf{varLag04}.
All that is required is to rewrite each
term appearing in the gauge variation (see \rf{appCC01},\rf{appCC02})
in terms of the operators
$\VV_{a,b}$.  We now represent
various pieces of the gauge variation of the vertices $\LL_\sma$.
To simplify our formulas, we adopt the following
convention. Let $A$ and $B$ be some operators. We use the relation
\be A \approx B \ee
in place of
\be e\psibr A \epsilonk =  e\psibr B \epsilonk + \hbox{ total
derivative}.\ee
Using such convention, we note that the gauge
variation of the minimal gravitational vertex, $\delta \LL_\smone$
\rf{appCC02}, can be read from the formula
\beq \label{M1FFdef01}
M_\smone (\alpha D+\FF) & \approx & \VV_{2,1}\,.
\eeq
From this formula, we see that the minimal gravitational vertex is
indeed not gauge invariant. In order to obtain gauge invariant
gravitational vertex, we should add to the minimal vertex,  two-derivative
vertices at least. Gauge variation of the two-derivative
vertices, $\delta\LL_\smtwo$ \rf{appCC02}, related to the vertex
operators $M_{2,a}$ can be read from the formulas%
\footnote{ The two-derivative vertex operators $M_{2,a}^\im$ are
not relevant in discussion of the gravitational vertex. These vertex operators
contribute only to the higher-derivative vertices of
the partial massless spin 5/2 field in $(A)dS$ space and the massless spin 5/2
field in flat space.
Contributions of $M_{2,a}^\im$ to the gauge variation is given below.}
\beq \label{Msmtwodef01}
\sum_{a=1}^3 M_{2,a}\, \alpha D & \approx & m_{2,1} \VV_{3,1} + (-4m_{2,1}(0) + 2
m_{2,2}) \VV_{3,2}
\nonumber\\
&+ & m_{2,3} \VV_{3,3}  +  m_{2,3} \VV_{3,4}
\nonumber\\[3pt]
& - & 2(m_1(0) + m_1(1))m_{2,1}(0) \VV_{2,1}
\nonumber\\
& + & (d -4) \fwt_3(0) m_{2,1}(0) \VV_{2,2}\,,
\\
\sum_{a=1}^3 M_{2,a} \, \FF  & \approx & \Bigl(\fwt_2(0)(4 m_{2,1}(0) + 2m_{2,2})
 + \fwt_1(0) m_{2,3}\Bigr) \VV_{2,1}
\nonumber \\
& + & \Bigl( \fwt_1(0)m_{2,1}(1) + \frac{3}{2}\fwt_2(0) m_{2,3}
\Bigr) \VV_{2,2}\,.
\eeq

Using these expressions, one can make sure that unwanted
$\VV_{2,1}$ term appearing in $\delta\LL_\smone +\delta\LL_\smtwo$
can be canceled by suitable choice of $m_{2,a}$. But, due to $\delta\LL_\smtwo$,
new terms governed by the operators $\VV_{2,2}$, $\VV_{3,a}$,
$a=1,2,3,4$, appear in
$\delta\LL_\smone +\delta\LL_\smtwo$. These new terms
should also be canceled to obtain gauge invariant Lagrangian $\LL_\smone
+\LL_\smtwo$.
One can make sure that this is not possible when $d>4$. This,
when $d>4$, the equation $\delta\LL_\smone +
\delta\LL_\smtwo=0$ can in no way be satisfied by choice of
$m_{2,a}$. However it turns out that the equation
$\delta\LL_\smone + \delta\LL_\smtwo=0$ has solution when $d=4$. Reason for this
is traced to the fact that the operators $\VV_{3,a}$, $a=1,2,3,4$,
are not all independent
when $d=4$. Namely, two of them, $\VV_{3,2}$ and $\VV_{3,4}$, are expressible in
terms of the remaining operators (see \rf{V32d4},\rf{V34d4}). Plugging
\rf{V32d4},\rf{V34d4} in \rf{Msmtwodef01}, one can make sure that
the equation $\delta\LL_\smone + \delta\LL_\smtwo=0$ can be satisfied by
suitable choice of $m_{2,a}$. Solution for such $m_{2,a}$ is given in
\rf{grav4d01sh},\rf{grav4d0501sh}.
This, to obtain the gravitational vertex for
$d>4$ we should add to $\LL_\smone$ and
$\LL_\smtwo$ the three-derivative vertices $\LL_\smthree$.
The gauge variation of these vertices can be read from the following
formulas:
\beq
M_{3,1}\alpha D & \approx &  - m_{3,1}(0)  \VV_{4,1} - \frac{1}{4}
\fwt_2 m_{3,1} \VV_{3,1}
\nonumber\\
& - & \fwt_3(0)  m_{3,1}(0) \VV_{3,3}
+ (d+3) \fwt_3(0)  m_{3,1}(0) \VV_{3,4}
\nonumber\\
& - & \frac{5}{4}\coscon m_{3,1}(0) \VV_{2,1} \,,
\\[10pt]
M_{3,2}\, \alpha D & \approx & m_{3,2} \VV_{4,1}
- \fwt_2(0) m_{3,2} \VV_{3,2}
\nonumber \\
&  + & 2\fwt_3(0)  m_{3,2} \VV_{3,3} - 2\fwt_3(0) m_{3,2} \VV_{3,4}
\nonumber\\
& + &    \Bigl( \frac{3}{4}(m_1^2(0) - m_1^2(1))  + \frac{10 -
5d}{4}\coscon \Bigr)  m_{3,2} \VV_{2,1}\,,
\nonumber\\
& - &  \frac{9}{2} \fwt_2(0) \fwt_3(0) m_{3,2} \VV_{2,2}\,,
\\[10pt]
M_{3,3}\alpha D & \approx &  - m_{3,3}(0) \VV_{4,2}  -  m_{3,3}(0)
\VV_{4,3}
\nonumber \\
& + & (d+2) \fwt_3(0)  m_{3,3}(0) N_\zeta \VV_{3,1}
+ d \fwt_3(1) m_{3,3}(0) \VV_{3,5}
\nonumber\\
&  - & \frac{3}{2}\coscon m_{3,3}(0) \VV_{2,2}\,,
\\[10pt]
M_{3,4}\alpha D & \approx &  \frac{1}{4}  m_{3,4}  \VV_{4,2}
+ \frac{1}{4}  m_{3,4} \VV_{4,3}
\nonumber\\
& - & 2 \fwt_3(0)   N_\zeta  m_{3,4} \VV_{3,1} - \frac{1}{2}(m_1(0) +
m_1(2)  )  m_{3,4} \VV_{3,3}
\nonumber\\
& - & m_1(1)  m_{3,4} \VV_{3,4}
- \fwt_3(1)  m_{3,4} \VV_{3,5}
\nonumber\\
& - & \frac{\coscon}{8}(d-7)  m_{3,4} \VV_{2,2}\,,
\\[10pt]
M_{3,5}\alpha D & \approx &  \frac{1}{2} \fwt_2(1) m_{3,5} \VV_{3,5}
\,,
\eeq

\beq
M_{3,1} \FF & \approx & \frac{3}{4}  \fwt_2 m_{3,1} \VV_{3,1}
-  \fwt_2(0) m_{3,1}(0) \VV_{3,2}
\nonumber\\
& + & 2 \fwt_3(0) m_{3,1}(0) \VV_{3,3}
+  \fwt_1(0) m_{3,1}(1) \VV_{3,4}\,,
\\[10pt]
M_{3,2} \FF & \approx & 4  \fwt_2(0) m_{3,2} \VV_{3,2} + 2(m_1(0) +
m_1(1)) \fwt_2(0) m_{3,2} \VV_{2,1}
\nonumber\\[5pt]
& + & 6 \fwt_2(0) \fwt_3(0) m_{3,2} \VV_{2,2}\,,
\\[10pt]
M_{3,3} \FF & \approx &  \Bigl(\fwt_1 (N_\zeta+1) m_{3,3} +2\fwt_3(0)
N_\zeta m_{3,3}(0) \Bigr) \VV_{3,1}
+ 4  \fwt_2(1) m_{3,3}(0) \VV_{3,3}
\nonumber\\
& + &  4  \fwt_2(0) m_{3,3}(0) \VV_{3,4}
+ \fwt_1(0)  m_{3,3}(1) \VV_{3,5} \,,
\\[10pt]
M_{3,4} \FF & \approx & \fwt_1(0) m_{3,4} \VV_{3,2} + \fwt_2(1)
m_{3,4} \VV_{3,3} + \fwt_2(0) m_{3,4} \VV_{3,4} \,,
\\[10pt]
\label{M35DDdef01} M_{3,5} \FF & \approx &
2 \fwt_1(1) m_{3,5} \VV_{3,3}
+ \frac{3}{2} \fwt_2(0) m_{3,5} \VV_{3,5}\,.
\eeq
Altogether, the relations in \rf{M1FFdef01}-\rf{M35DDdef01}
give the gauge variation in \rf{varLag}-\rf{varLag04} with
the $X_{a,b}$ defined in
\rf{X41def}-\rf{X22def01}.
We note that, due to $ \delta\LL_\smthree $, new terms governed by the vertex
operators $\VV_{3,5}$, $\VV_{4,a}$, $a=1,2,3$, appear in
the gauge variation $\delta(\LL_\smone +\LL_\smtwo + \LL_\smthree)$.
Now, the equation $\delta(\LL_\smone +\LL_\smtwo + \LL_\smthree) =0$
can be solved for arbitrary $d>4$
by suitable choice of the coefficients $m_{3,a}$ and $m_{2,a}$.
To summarize, when $d>4$, the procedure of building
the gravitational vertex of the massive
spin 5/2 field terminates at three-derivative vertices.

The remaining two-derivative and three derivative vertex
operators $M_{2,a}^\im$ and
$M_{3,a}^\im$ do not contribute to the gravitational vertex. But, these operators
lead to some higher-derivative vertices. The gauge variation of
vertices associated with the operators $M_{2,a}^\im$,
$M_{3,a}^\im$ can be read from the following formulas:
\beq \label{M23gauvarcomdef01}
M_{2,3}^\im \alpha D & \approx & m_{2,3}^\im \VV_{3,3} -  m_{2,3}^\im
\VV_{3,4}\,,
\\[3pt]
M_{2,3}^\im \FF & \approx & \fwt_1(0) m_{2,3}^\im \VV_{2,1} -
\frac{3}{2}\fwt_2(0) m_{2,3}^\im \VV_{2,2}\,,
\eeq
\beq
M_{3,3}^\im\alpha D & \approx &  - m_{3,3}^\im(0) \VV_{4,2}  +
m_{3,3}^\im(0) \VV_{4,3}
\nonumber \\[3pt]
& + & (d+2) \fwt_3(0)  m_{3,3}^\im(0) N_\zeta \VV_{3,1}
- d \fwt_3(1) m_{3,3}^\im(0) \VV_{3,5}
\nonumber\\[3pt]
&  + & \frac{3}{2}\rho m_{3,3}^\im(0) \VV_{2,2}\,,
\\[7pt]
M_{3,4}^\im\alpha D & \approx &  \frac{1}{4}  m_{3,4}^\im  \VV_{4,2}
- \frac{1}{4}  m_{3,4}^\im \VV_{4,3}
\nonumber\\[3pt]
& - &  \fwt_3(0)   N_\zeta  m_{3,4}^\im \VV_{3,1} - \frac{1}{2}(m_1(0)
+ m_1(2)  )  m_{3,4}^\im \VV_{3,3}
\nonumber\\[3pt]
& + & m_1(1)  m_{3,4}^\im \VV_{3,4}
+ \fwt_3(1)  m_{3,4}^\im \VV_{3,5}
\nonumber\\[3pt]
& + & \frac{\rho}{8}(d-7)  m_{3,4}^\im \VV_{2,2}\,,
\\[7pt]
M_{3,5}^\im\alpha D & \approx &  - \frac{1}{2} \fwt_2(1) m_{3,5}^\im
\VV_{3,5} \,,
\eeq
\beq
M_{3,3}^\im \FF & \approx &  \Bigl( (N_\zeta+1) \fwt_1  m_{3,3}^\im -
2\fwt_3(0) N_\zeta m_{3,3}^\im(0) \Bigr) \VV_{3,1}
+ 4  \fwt_2(1) m_{3,3}^\im(0) \VV_{3,3}
\nonumber\\[5pt]
& - &  4  \fwt_2(0) m_{3,3}^\im(0) \VV_{3,4}
- \fwt_1(0)  m_{3,3}^\im(1) \VV_{3,5} \,,
\\[10pt]
M_{3,4}^\im \FF & \approx
 & \fwt_1(0) m_{3,4}^\im \VV_{3,2} + \fwt_2(1) m_{3,4}^\im
\VV_{3,3} - \fwt_2(0) m_{3,4}^\im \VV_{3,4} \,,
\\[10pt]
\label{M35comdef} M_{3,5}^\im \FF & \approx &
2 \fwt_1(1) m_{3,5}^\im \VV_{3,3}
- \frac{3}{2} \fwt_2(0) m_{3,5}^\im \VV_{3,5}\,.
\eeq
Altogether, the relations in \rf{M23gauvarcomdef01}-\rf{M35comdef}
give $X_{a,b}^\im$ terms in
gauge variation \rf{varLag}-\rf{varLag04} with the $X_{a,b}^\im$ defined in
\rf{X43comdef01}-\rf{X22cdef01}.

Up to this point we have considered the case of $d>4$. We now explain how the
results above discussed  can be used for derivation of $X_{a,b}$, $X_{a,b}^\im$ in
$d=4$.

Firstly, we note that, when $d=4$,
the operators $M_{3,1}$, $M_{3,2}$, $M_{3,4}$, $M_{3,4}^\im$
are expressible in terms of the remaining operators $M_{a,b}$, $M_{a,b}^\im$.
This can be taken into account by equating to zero the corresponding
coefficients
$m_{3,1}$, $m_{3,2}$, $m_{3,4}$, $m_{3,4}^\im$ (see \rf{mabzeroval01})
in the expressions for $X_{a,b}$, $X_{a,b}^\im$ given in \rf{X41def}-\rf{X22cdef01}.
In this way, we obtain that $X_{4,1}$ (see \rf{X41def})
is automatically equal to zero, while
the equations $X_{4,a}=0$, $X_{4,a}^\im=0$, $a=2,3$,
(see \rf{X4223def},\rf{X43comdef01}) give
\be\label{m33m33imsol} m_{3,3}(0)=0\,,\qquad \quad m_{3,3}^\im(0)=0\,.\ee

Secondly, when $d=4$,  the operators $\VV_{3,2}$ and $\VV_{3,4}$,
are expressible in terms of
the remaining operators $\VV_{3,a}, $$\VV_{2,a}$.
This, when $d=4$, basis of the operators $\VV_{a,b}$ is given by
\rf{veropebasd4nn}. To take this into account, we substitute
the representation for
the operators $\VV_{3,2}$ and $\VV_{3,4}$ given in \rf{V32d4},\rf{V34d4}
into the gauge variation (see \rf{varLag02}-\rf{varLag04})
and represent the gauge variation in terms of base operators $\VV_{a,b}$
\rf{veropebasd4nn}. In this way, we transform the gauge variation to the form
\rf{varLag02d4}-\rf{varLag04d4} and, using \rf{m33m33imsol}, we
obtain the corresponding
$X_{a,b}$, $X_{a,b}^\im$,
\beq \label{X31def01nnn}
X_{3,1} & =& \frac{1}{2}(1-N_\zeta)(4m_{2,1}(0) - m_{2,2}) +
N_\zeta (m_{2,1}(1) + 2\fwt_1(1) m_{3,3}(1))\,,
\\[3pt]
X_{3,3} & = &  2\fwt_1(1) m_{3,5} + m_{2,3}\,,
\\[3pt]
X_{3,5} & =& \fwt_1(0) m_{3,3}(1) + \frac{3}{4} \mas m_{3,5}\,,
\\[3pt]
X_{2,1} & = &
-2\mas m_{2,1}(0) - \half \mas  m_{2,2} + \fwt_1(0) m_{2,3} +
\sqrt{2}\kappa\,,
\\[3pt]
\label{X22def01nnn} X_{2,2} & = & \fwt_1(0) m_{2,1}(1) +
\frac{3}{4}\mas m_{2,3}\,,
\eeq
\beq
&& X_{3,1}^\im = 2N_\zeta\fwt_1(1) m_{3,3}^\im(1) \,,
\\[3pt]
&&  X_{3,3}^\im = m_{2,3}^\im
+ 2\fwt_1(1) m_{3,5}^\im\,,
\\[3pt]
&& X_{3,5}^\im =
- \fwt_1(0) m_{3,3}^\im(1) - \frac{3}{4} \mas m_{3,5}^\im\,,
\\[3pt]
&&  X_{2,1}^\im =  \fwt_1(0) m_{2,3}^\im  \,,
\\[3pt]
\label{X22d401}
&& X_{2,2}^\im = -  \frac{3}{4} \mas m_{2,3}^\im\,.
\eeq
All solutions to the equations $X_{a,b}=0$, $X_{a,b}^\im=0$ were discussed in Sections
\ref{secmasverd401}, \ref{speintver01}. From \rf{X31def01nnn}-\rf{X22d401},
it is easy to see that solution to the equations for the gravitational vertex, $X_{a,b}=0$,
$X_{a,b}^\im=0$, $\kappa \ne 0$,
is indeed given by \rf{grav4d01sh},\rf{grav4d0501sh}. In other words, in $d=4$,
the gravitational vertex does not involve three-derivative contributions.

We now outline procedure for deriving relations
\rf{M1FFdef01}-\rf{M35comdef}.
Let $X$ be operator constructed out the Weyl tensor, the oscillators,
and the covariant derivative. Then, we use the following procedure.
\\
{\bf i}) Expressions like $\psibr X \alpha D\epsilonk $ and
$\psibr\bar\alpha D X \epsilonk$ can be rewritten as $\psibr [X,
\alpha D] +\alpha D X \epsilonk $ and $\psibr [\bar\alpha D, X] +
X\bar\alpha D \epsilonk$ respectively. Expressions like  $\psibr [X,
\alpha D]\epsilonk $ and $\psibr[\bar\alpha D, X]\epsilonk$ can be
evaluated by using commutators for the covariant derivative and various
relations for the Weyl tensor. The remaining expressions like $\psibr
\alpha D X \epsilonk  $ and $\psibr X \bar\alpha D \epsilonk$ are
evaluated by using Lorentz constraints  for $\psibr$ \rf{Lorconst}
and $\epsilonk$ \rf{Lorconeps} to obtain
\beq
&&  \alpha D X \approx  \zeta (d+ 2-2N_\zeta) \fwt_3 X  \,,
\\[3pt]
&& X \bar\alpha D \approx   - d \fwt_3(0) X \bar\zeta \,.
\eeq
{\bf ii}) In expressions like $\psibr\Dline X \epsilonk$ and $\psibr X
\Dline \epsilonk$, use E.o.M for $\psibr$ and $\epsilonk$ to obtain
the relations
\beq
&& \Dline X \approx  (-  m_1 + 2\zeta \fwt_3 \gamma\bar\alpha)X \,,
\\[3pt]
&& X \Dline \approx X ( - m_1^\smone - 2\gamma\alpha \fwt_3(0)
\bar\zeta)\,.
\eeq
{\bf iii}) In expressions like $\psibr X \gamma\alpha \epsilonk $, (anti)commute
$\gamma\alpha$ to left, while, in expressions like $\psibr
\gamma\bar\alpha X \epsilonk$, (anti)commute $\gamma\bar\alpha$ to right, and
then use the $\gamma$-tracelessness constraints to obtain the relations
\be
\gamma \alpha X \approx  0   \,, \qquad\qquad
X \gamma\bar\alpha \approx 0 \,.
\ee
{\bf iv})  Rewrite expressions like $\psibr [D^A, X] D^A \epsilonk $ as
\be   [D^A,X] D^A \approx \frac{1}{2} D^2 X  -
\frac{1}{2} X D^2  - \frac{1}{2}  [D^A,[D^A,X]]
\,,\ee
$D^2\equiv D^AD^A$, and then use the relations
\beq
&& \hspace{-1.8cm} X \Box   \approx  X \Bigl( m_1^\smone m_1^\smone
+ \rho (\frac{d(d-1)+4}{4}- N_\zeta)
-  4\gamma\alpha \fwt_2(1) \fwt_3(0) \bar\zeta - 4\alpha D \fwt_3(0) \bar\zeta
\Bigr)\,,
\\[5pt]
&& \hspace{-1.8cm}
\Box X \approx  \Bigl( m_1^2 + \rho (\frac{d(d-1)+8}{4}- N_\zeta)
+  4 \zeta \fwt_2\fwt_3\gamma\bar\alpha - 4 \zeta^2
\fwt_3(0)\fwt_3(1) \bar\alpha^2 + 4 \zeta \fwt_3 \bar\alpha D \Bigr)X\,,
\eeq
which are obtainable from E.o.M for $\psibr$ and $\epsilonk$.
The operator $\Box$ is
defined in \rf{Dcovdef01}.
Expression like $[D^A,[D^A, X]]$ can be simplified by using
relations for the Weyl tensor given in
\rf{DDCeq}-\rf{trbia02}.

\appendix{ Basis of vertex operators in $d>4$}

For $d>4$, the vertex operators defined in \rf{M21def}-\rf{M35cdef}
constitute a complete basis of on-shell two-derivative and three-derivative
vertex operators. This is to say that, on-shell,
all remaining two-derivative and three-derivative
vertex operators constructed out the Weyl tensor and covariant derivative
can be expressed in terms of operators \rf{M21def}-\rf{M35cdef}.
Note that vertex operators $A$ and $B$ are considered to be equivalent
if they satisfy the relation
\be \label{simconven02}  e\psibr A \psik  =  e\psibr B \psik + \hbox{ total
derivative},\ee
which we simply rewrite as
\be \label{simconven01} A \sim B \,.\ee
To discuss basis of vertex operators in $d>4$, we drop
the coefficients $m_{a,b}$, $m_{a,b}^\im$  in \rf{M21def}-\rf{M35cdef}
and obtain the following
{\bf basis of two-derivative and three-derivative vertex operators for $d>4$}:
\beq
\label{MMbasdef01} &&  \hspace{0cm} \MM_{2,1}(a) = C^{ABCE} \gamma^{AB}
\alpha^C  \bar\alpha^E \fwh(a)\,, \qquad a=0,1\,,
\\[3pt]
&&\label{MMbasdef02}  \hspace{0cm}  \MM_{2,2} =  C^{ABCE} \alpha^A\alpha^E
\bar\alpha^B\bar\alpha^C \,,
\\[3pt]
&&\label{MMbasdef03}  \hspace{0cm}   \MM_{2,3} =  C^{ABCE}
\gamma^A \left( \alpha^B
\alpha^C \bar\alpha^E \bar\zeta
- \zeta \alpha^E \bar\alpha^B \bar\alpha^C \right)\,,
\\[15pt]
&&\label{MMbasdef04}   \MM_{3,1}(a) =  C^{A(BC)E} \gamma^A
\alpha^B \bar\alpha^C D^E\fwh(a)\,,\qquad a=0,1\,,
\\[3pt]
\label{MMbasdef05} && \MM_{3,2} = \frac{1}{2} \DD^F C^{ABCE} \gamma^A
(\alpha^B \alpha^C \bar\alpha^E\bar\alpha^F -
\alpha^E\alpha^F\bar\alpha^B \bar\alpha^C)\,,
\\[3pt]
\label{MMbasdef06} && \MM_{3,3}(a) = C^{ABCE} \gamma^{AB}
\left(\alpha^C \fwh(a) \bar\zeta
+  \zeta \fwh(a) \bar\alpha^C  \right) D^E\,,\qquad a=0,1\,,
\\[3pt]
\label{MMbasdef07} && \MM_{3,4} =  C^{ABCE} \left( \alpha^B\alpha^C
\bar\alpha^A \bar\zeta
- \zeta \alpha^A \bar\alpha^B \bar\alpha^C \right) D^E\,,
\\[3pt]
\label{MMbasdef08} && \MM_{3,5} = C^{ABCE} \gamma^A \left( \alpha^B
\alpha^C \bar\zeta^2
+ \zeta^2 \bar\alpha^B \bar\alpha^C \right) D^E\,,
\\[5pt]
\label{MMbasdef09} &&\hspace{1cm}
\MM_{2,3}^\im =  C^{ABCE} \gamma^A \left( \alpha^B
\alpha^C \bar\alpha^E \bar\zeta
+ \zeta \alpha^E \bar\alpha^B \bar\alpha^C \right)\,,
\\[5pt]
\label{MMbasdef10} && \hspace{1cm}
\MM_{3,3}^\im(a) = C^{ABCE} \gamma^{AB}
\left(\alpha^C \fwh(a) \bar\zeta
-  \zeta \fwh(a) \bar\alpha^C  \right) D^E\,,\qquad a=0,1\,,
\\[5pt]
\label{MMbasdef11} && \hspace{1cm}
\MM_{3,4}^\im =  C^{ABCE} \left( \alpha^B\alpha^C
\bar\alpha^A \bar\zeta
+ \zeta \alpha^A \bar\alpha^B \bar\alpha^C \right) D^E\,,
\\[5pt]
\label{MMbasdef12} && \hspace{1cm}
\MM_{3,5}^\im = C^{ABCE} \gamma^A \left( \alpha^B
\alpha^C \bar\zeta^2
- \zeta^2 \bar\alpha^B \bar\alpha^C \right) D^E\,,
\eeq
where operators $\MM(a)$,
\rf{MMbasdef01},\rf{MMbasdef04},\rf{MMbasdef06},\rf{MMbasdef10},
depend on the operator $N_\zeta$
through the function $\fwh(a)$ defined by
\be \label{fwhdef01} \fwh(0) = 1-N_\zeta\,,\qquad \fwh(1)= N_\zeta \,. \ee
This implies that each operator $\MM(a)$ provides
two base operators: $\MM(0)$ and $\MM(1)$.
The representation for operators $\MM = \MM_{a,b}, \MM_{a,b}^\im$
given in \rf{MMbasdef01}-\rf{MMbasdef12} is obtained
by requiring that:

\noindent {\bf i}) $\MM$ do not involve higher than second order terms
in $\alpha^A$, $\zeta$, $\bar\alpha^A$, $\bar\zeta$ and satisfy the
commutators $[N_\alpha +N_\zeta,\MM]=0$;

\noindent {\bf ii}) $\MM$ do not involve terms like
$\gamma\alpha p_1 $ and $p_2\gamma\bar\alpha$,
where $p_{1,2}$ are polynomial in the oscillators (such terms in view
of the $\gamma$-tracelessness  constraint do not contribute to $\LL_\sma$).

We note that under the hermitian conjugation
the operators $\MM_{a,b}, \MM_{a,b}^\im$ transform as
\be \label{herconrul01}
(e\MM_{a,b})^\dagger = -\gamma^0 e \MM_{a,b} \gamma^0\,,
\qquad  (e\MM_{a,b}^\im)^\dagger  = + \gamma^0 e \MM_{a,b}^\im  \gamma^0\,,\ee
$e\equiv \det e_\mu^A$. To derive this, we use \rf{DDCeq2}.
Relations \rf{herconrul01} imply that in order for the interaction vertices
$\LL_\smtwo$, $\LL_\smthree$ \rf{LLaMMadef01} to be real-valued,
the operators $\MM_{a,b}$ and $\MM_{a,b}^\im$
should be multiplied by the respective real-valued coefficients $m_{a,b}$
and purely-imagine coefficients $m_{a,b}^\im$ (see \rf{M21def}-\rf{M35cdef},
\rf{mabreacon01},\rf{mabreacon02}).

To demonstrate that $\MM_{a,b}$, $\MM_{a,b}^\im$
form a basis of vertex operators and
to illustrate the techniques, let us consider the following
vertex operator $C^{ABCE} \gamma^A\alpha^C \bar\alpha^E D^B \fwh(a)$.
This vertex operator does not enter
into our vertices basis, and hence it should be expressible
in terms of $\MM_{a,b}$
and $\MM_{a,b}^\im$.
To show that this in fact the case, we prove the relation
\be
\label{MMbasdef26n1} C^{ABCE} \gamma^A\alpha^C \bar\alpha^E D^B \fwh(a)
\sim - 2\fwt_3(0) \delta_{a,0} \MM_{2,3}^\im\,,
\ee
where $\delta_{a,0}= 1(0)$  for $a=0 (1)$.
To this end, we firstly note the relations
\beq \label{s100-0101n1n05n1}
&& \MM_{2,1}(a) \Dline \sim  - m_1  \MM_{2,1}(a) - 8 \fwt_3(0) \delta_{a,0}
\MM_{2,3}^+\,,
\\[3pt]
\label{s100-0101n1n10n1}
&& \Dline \MM_{2,1}(a) \sim -m_1 \MM_{2,1}(a) + 8 \fwt_3(0) \delta_{a,0}
\MM_{2,3}^-\,,
\eeq
where $\MM_{2,3}^\pm$ are defined in \rf{MMbasdef04n},\rf{MMbasdef05n}.
To prove \rf{s100-0101n1n05n1}, we use E.o.M for $\psik$, \rf{sec5006},
and obvious formula
\be \label{s100-0101n1n09}
\alpha^A \alpha^B \bar\alpha^C \fwh(a) \bar\zeta
\sim \delta_{a,0} \alpha^A \alpha^B \bar\alpha^C \bar\zeta\,,
\ee
which just reflects the fact that $\psik$ is a
degree-2 homogeneous polynomial in the oscillators $\alpha^A$, $\zeta$.
Proof of relation \rf{s100-0101n1n10n1} is similar:
we use E.o.M for $\psibr$ and obvious formula
\be
\zeta \alpha^A \bar\alpha^B \bar\alpha^C \fwh(a)
\sim \delta_{a,0} \zeta \alpha^A \bar\alpha^B \bar\alpha^C\,.  \ee

Secondly,
taking commutators of relations \rf{s100-0101n1n05n1}, \rf{s100-0101n1n10n1},
we obtain
\be \label{s100-0101n1n11n1}
[\MM_{2,1}(a),\Dline] \sim - 8 \fwt_3(0) \delta_{a,0} \MM_{2,3}^\im\,.\ee
Then, using the well-known commutator
\be \half [\gamma^{AB},\gamma^C]
= \gamma^A\eta^{BC} - \gamma^B\eta^{AC}\,,
\ee
and relations for the Weyl tensor given in
\rf{DDCeq}-\rf{trbia02},
we rewrite l.h.s of \rf{s100-0101n1n11n1}
as
\be \label{s100-0101n1n12n1}
[\MM_{2,1}(a),\Dline] \sim 4
C^{ABCE} \gamma^A\alpha^C \bar\alpha^E D^B \fwh(a)\,. \ee
Comparing \rf{s100-0101n1n11n1} with \rf{s100-0101n1n12n1}
gives desired relation \rf{MMbasdef26n1}.

\appendix{ Basis of vertex operators in
$d=4$}

For $d=4$, the basis of vertex operators
\rf{M21def}-\rf{M35cdef} turns out to be over-completed, i.e.,
some of the vertex operators in
\rf{M21def}-\rf{M35cdef} (or equivalently \rf{MMbasdef01}-\rf{MMbasdef12})
can be expressed in terms of others.
This is to say that, using convention in \rf{simconven01},\rf{simconven02},
we have the following relations for the vertex
operators when $d=4$:
\beq
\label{sec100-01} &&  \MM_{3,1}(a) \,\,\sim\,\,  \frac{1}{4}m_1
\MM_{2,1}(a) - \frac{1}{4} \fwt_1(0) \MM_{2,3} \delta_{a,0}\,, \qquad
\quad a=0,1\,,
\\[3pt]
\label{sec100-02} &&  \MM_{3,2} \,\,\sim\,\, - \frac{1}{2}
m_1 \MM_{2,1}(0) - \mas \MM_{2,2} -\half  \fwt_1(0)  \MM_{2,3}\,,
\\[3pt]
\label{sec100-01n001}
&&  \MM_{3,4} \,\,\sim\,\,
- \frac{1}{4}\MM_{3,3}(0)   -\frac{1}{2} \mas \MM_{2,3}\,,
\\[3pt]
\label{sec100-01n002}
&& \MM_{3,4}^\im \,\,\sim\,\,  - \frac{1}{4} \MM_{3,3}^\im(0)
- \frac{1}{2} \mas \MM_{2,3}^\im\,,
\eeq
where $\delta_{a,0}= 1(0)$  for $a=0 (1)$.
Taking these relations into account we find the following
\\
{\bf basis of
two-derivative and three-derivative vertex operators in $d=4$}:
\beq
\label{MMbasdef01d4} && \MM_{2,1}(a) = C^{ABCE} \gamma^{AB}
\alpha^C \bar\alpha^E \fwh(a)\,, \qquad\quad a =0,1\,,
\\[5pt]
&&\label{MMbasdef02d4} \MM_{2,2} =  C^{ABCE} \alpha^A\alpha^E
\bar\alpha^B\bar\alpha^C \,,
\\[5pt]
&&\label{MMbasdef03d4} \MM_{2,3} =  C^{ABCE} \gamma^A \left( \alpha^B
\alpha^C \bar\alpha^E \bar\zeta
- \zeta \alpha^E \bar\alpha^B \bar\alpha^C \right)\,,
\\[5pt]
\label{MMbasdef06d4} && \MM_{3,3}(a) = C^{ABCE} \gamma^{AB}
\left(\alpha^C \fwh(a) \bar\zeta
+  \zeta \fwh(a) \bar\alpha^C  \right) D^E\,, \qquad\ a=0,1\,,
\\[5pt]
\label{MMbasdef08d4} && \MM_{3,5} = C^{ABCE} \gamma^A \left( \alpha^B
\alpha^C \bar\zeta^2
+ \zeta^2 \bar\alpha^B \bar\alpha^C \right) D^E\,,
\\[5pt]
&&\hspace{1cm}
\label{MMbasdef09d4} \MM_{2,3}^\im =  C^{ABCE} \gamma^A \left(
\alpha^B \alpha^C \bar\alpha^E \bar\zeta
+ \zeta \alpha^E \bar\alpha^B \bar\alpha^C \right)\,,
\\[5pt]
\label{MMbasdef10d4} && \hspace{1cm}
 \MM_{3,3}^\im(a) = C^{ABCE} \gamma^{AB}
\left(\alpha^C \fwh(a) \bar\zeta
-  \zeta \fwh(a) \bar\alpha^C  \right) D^E\,, \qquad\ a=0,1\,,
\\[5pt]
\label{MMbasdef12d4} && \hspace{1cm}
\MM_{3,5}^\im = C^{ABCE} \gamma^A \left(
\alpha^B \alpha^C \bar\zeta^2
- \zeta^2 \bar\alpha^B \bar\alpha^C \right) D^E\,,
\eeq
where $\fwh(a)$ is as shown in \rf{fwhdef01}.
To illustrate the techniques, we derive
relation \rf{sec100-01}. We start with the identity
\be\label{subsec100-0101n1} C^{AB[CE}\, \gamma^{ABF]} \alpha^C
\bar\alpha^E D^F \fwh(a) =0\,, \ee
which reflects the fact that antisymmetrization w.r.t five
indices $ABCEF$ gives zero
when $d=4$. In \rf{subsec100-0101n1} and below,
antisymmetrization of three $\gamma$ matrices in $\gamma^{ABC}$
is normalized so that
$\gamma^{012}=\gamma^0\gamma^1\gamma^2$.
Rewriting identity \rf{subsec100-0101n1} as
\be \label{subsec100-0101n1n01}
\Bigl( C^{ABCE}\gamma^{ABF}\alpha^C \bar\alpha^E D^F
+
C^{ABFC}\gamma^{ABE}\alpha^C \bar\alpha^E D^F
+ C^{ABEF}\gamma^{ABC}\alpha^C \bar\alpha^E D^F\Bigr)\fwh(a)  = 0\,,
\ee
we note that the first term and sum of the second and third terms
in \rf{subsec100-0101n1n01} can be rewritten as
\beq  \label{subsec100-0101n1n02}
&& C^{ABCE}\gamma^{ABF}\alpha^C \bar\alpha^E D^F \fwh(a) =
\MM_{2,1}(a)\Dline - 2C^{ABCE}\gamma^A \alpha^C\bar\alpha^E D^B\,,
\\[5pt]
\label{subsec100-0101n1n03}
&& C^{ABFC}\gamma^{ABE}\alpha^C \bar\alpha^E D^F \fwh(a)
+ C^{ABEF}\gamma^{ABC}\alpha^C \bar\alpha^E D^F\fwh(a)  \sim 4 \MM_{3,1}(a)
\eeq
respectively. To derive \rf{subsec100-0101n1n02}, we just use the obvious relation
\be
\gamma^{ABF} = \gamma^{AB}\gamma^F - \gamma^A\eta^{BF}
+ \gamma^B \eta^{AF}\,,
\ee
while to derive \rf{subsec100-0101n1n03} we use the relations
\beq
&& \gamma^{ABE} = \gamma^{AB} \gamma^E - \gamma^A\eta^{BE}
+ \gamma^B \eta^{AE}\,,
\\
&& \gamma^{ABC} = \gamma^C \gamma^{AB}- \eta^{CA}\gamma^B
+ \eta^{CB}\gamma^A\,,
\eeq
and the $\gamma$-tracelessness condition for $\psibr$.
Taking relations \rf{subsec100-0101n1n02},
\rf{subsec100-0101n1n03} into account we see that identity
\rf{subsec100-0101n1n01}
amounts to the relation
\beq \label{s100-0101n1n04}
&&
\MM_{2,1}(a)\Dline - 2C^{ABCE}\gamma^A\alpha^C\bar\alpha^E D^B
+ 4 \MM_{3,1}(a) \sim 0\,. \eeq
In view of \rf{MMbasdef26n1},\rf{s100-0101n1n05n1}, which are valid for
$d\geq 4$, and the relation $\fwt_1(0)=-4\fwt_3(0)$,
we see that \rf{s100-0101n1n04} amounts to
relation \rf{sec100-01}.

In similar way, one can prove the remaining relations
\rf{sec100-02}-\rf{sec100-01n002}. To this end, we use the following respective
identities:
\beq\label{identN301n1} && C^{AB[CE}\,\alpha^{|E|} \alpha^A \bar\alpha^B
\gamma^{F]}\, \bar\alpha^C D^F =0\,,
\\[3pt]
\label{subsec1000301n1} && C^{AB[CE}\,\gamma^{AF}\,\alpha^{B]}
\alpha^C\bar\alpha^E \bar\zeta D^F =0\,,
\\[3pt]
\label{sbsc1004-0101n1} && C^{AB[CE} \zeta \alpha^{|E|} \gamma^{AFB]}
\bar\alpha^C D^F =0\,.
\eeq
We note that \rf{identN301n1} leads to \rf{sec100-02}, while
\rf{identN301n1},\rf{subsec1000301n1} lead
to
\beq
&& \label{sec100-03} \MM_{3,4}^+ = - \frac{1}{4}
\MM_{3,3}^+(0) -\frac{1}{2} \mas \MM_{2,3}^+\,,
\\[3pt]
&& \label{sec100-03n1} \MM_{3,4}^- =  \frac{1}{4} \MM_{3,3}^-(0)
-\frac{1}{2} \mas \MM_{2,3}^-\,,
\eeq
where we use the notation
\beq
&&
\label{MMbasdef04n} \MM_{2,3}^+ =  C^{ABCE} \gamma^A  \alpha^B
\alpha^C \bar\alpha^E \bar\zeta\,,
\\[3pt]
&& \label{MMbasdef05n} \MM_{2,3}^- =  C^{ABCE} \gamma^A  \zeta
\alpha^E \bar\alpha^B \bar\alpha^C\,,
\\[3pt]
\label{MMbasdef13} && \MM_{3,3}^+ (a)= C^{ABCE} \gamma^{AB} \alpha^C
\fwh(a) \bar\zeta D^E\,,
\\[3pt]
\label{MMbasdef14} && \MM_{3,3}^-(a) = C^{ABCE} \gamma^{AB} \zeta \fwh(a) \bar
\alpha^C D^E\,,
\\[3pt]
\label{MMbasdef15} && \MM_{3,4}^+ = C^{ABCE} \alpha^B\alpha^C
\bar\alpha^A \bar\zeta  D^E\,,
\\[3pt]
\label{MMbasdef16} && \MM_{3,4}^- = C^{ABCE} \zeta \alpha^A
\bar\alpha^B \bar\alpha^C  D^E\,.
\eeq
In view of relations
\beq
\label{MMbasdef19} && \MM_{2,3}  = \MM_{2,3}^+ - \MM_{2,3}^-\,,
\\[3pt]
\label{MMbasdef20} && \MM_{3,3}(a) = \MM_{3,3}^+(a)  +  \MM_{3,3}^-(a)\,,
\\[3pt]
\label{MMbasdef21} && \MM_{3,4} = \MM_{3,4}^+ - \MM_{3,4}^-\,,
\\[10pt]
\label{MMbasdef19n} && \MM_{2,3}^\im = \MM_{2,3}^+ + \MM_{2,3}^-\,,
\\[3pt]
\label{MMbasdef20n} && \MM_{3,3}^\im(a) = \MM_{3,3}^+(a) - \MM_{3,3}^-(a)\,,
\\[3pt]
\label{MMbasdef21n} && \MM_{3,4}^\im = \MM_{3,4}^+ + \MM_{3,4}^-\,,
\eeq
we see that \rf{sec100-03},\rf{sec100-03n1}
amount to \rf{sec100-01n001},\rf{sec100-01n002}.

To finish discussion in this Appendix, we note that,
when $d=4$, basis of the operators $\VV_{a,b}$
\rf{VV41def01}-\rf{VV22def01} also
turns out to be over-completed, i.e.,
some of the operators in
\rf{VV41def01}-\rf{VV22def01} can be expressed in terms of others. Namely,
when $d=4$, the following relations hold true
\beq
\label{V32d4} &&  \VV_{3,2} \approx \frac{1}{4}(N_\zeta-1) \VV_{3,1}
-\frac{1}{2} \mas \VV_{2,1}\,,
\\
\label{V34d4} &&  \VV_{3,4} \approx \frac{3}{8} \mas \VV_{2,2}\,.
\eeq
These relations can be obtained by using the following respective
identities:
\beq
&& C^{AB[CE} \gamma^{AF}\alpha^{B]} \alpha^C \bar \alpha^E D^F
 =0 \,,
\\
&& C^{AB[CE} \gamma^{ABF]} \zeta \alpha^C \bar \alpha^E D^F
 =0 \,.
\eeq

\end{document}